\documentclass{article}

\usepackage{aas_macros}

\usepackage[english]{babel}

\usepackage[letterpaper,top=2cm,bottom=2cm,left=3cm,right=3cm,marginparwidth=1.75cm]{geometry}

\usepackage{amsmath}
\usepackage{graphicx}
\usepackage[colorlinks=true, allcolors=blue]{hyperref}
\usepackage{authblk}

\usepackage{natbib}

\usepackage{subfigure}
\usepackage{amsmath}

\date{August 1, 2022}

\begin{document}
\title{Determining the dust environment of an unknown comet for a spacecraft fly-by: The case of ESA's Comet Interceptor mission.}

\author[1,2]{Raphael Marschall}
\author[3,4]{Vladimir Zakharov}
\author[3]{Cecilia Tubiana}
\author[5]{Michael S. P. Kelley}
\author[6]{Carlos Corral van Damme}
\author[7]{Colin Snodgrass}
\author[8,9]{Geraint H. Jones}
\author[10]{Stavro L. Ivanovski}
\author[11]{Frank Postberg}
\author[3]{Vincenzo Della Corte}
\author[12]{Jean-Baptiste Vincent}
\author[13]{Olga Muñoz}
\author[14]{Fiorangela La Forgia}
\author[15]{Anny-Chantal Levasseur-Regourd}
\author{and the Comet Interceptor Team}

\affil[1]{CNRS, Observatoire de la Côte d'Azur, Laboratoire J.-L. Lagrange, CS 34229, 06304 Nice Cedex 4, France; raphael.marschall@oca.eu}
\affil[2]{Southwest Research Institute, 1050 Walnut St, Suite 300, Boulder, CO 80302, USA}
\affil[3]{INAF – Istituto di Astrofisica e Planetologia Spaziali, Area Ricerca Tor Vergata, Via Fosso del Cavaliere 100, 00133 Rome, Italy}
\affil[4]{LESIA, Observatoire de Paris, Université PSL, CNRS, Sorbonne Universite, Universite de Paris, 5 Place Jules Janssen, 92195 Meudon, France}
\affil[5]{Department of Astronomy, University of Maryland, College Park, MD 20742-0001, USA}
\affil[6]{European Space Research and Technology Centre (ESTEC), European  Space  Agency  (ESA),  Keplerlaan  1,  2201-AZ  Noordwijk, The Netherlands}
\affil[7]{Institute for Astronomy, University of Edinburgh, Royal Observatory, Edinburgh, EH9 3HJ, UK}
\affil[8]{Mullard Space Science Laboratory, University College London, Holmbury St. Mary, Dorking, Surrey, RH5 6NT, UK}
\affil[9]{The Centre for Planetary Sciences at UCL/Birkbeck, London WC1E 6BT, UK}
\affil[10]{INAF- Osservatorio Astronomico di Trieste, Italy}
\affil[11]{Institute of Geological Sciences, Freie Universität Berlin, Germany}
\affil[12]{DLR Institute of Planetary Research, Germany}
\affil[13]{Instituto de Astrofisica de Andalucia, CSIC. Glorieta de la Astronomia sn. Granada 18008, Spain}
\affil[14]{Department of Physics and Astronomy, University of Padova, vicolo Osservatorio, 3 Padova, Italy}
\affil[15]{LATMOS, Sorbonne Univ., CNRS, Campus Pierre et Marie Curie, BC 102, 4 place Jussieu, 75005 Paris, France}

\maketitle

\begin{abstract}
   {\noindent\textbf{Context:} Assessment of the dust environment of a comet is needed for data analysis and planning of spacecraft missions, such as ESA's Comet Interceptor (CI) mission. The distinctive feature of CI is that the target object will be defined shortly before (or even after) launch therefore the properties of the nucleus and dust environment are poorly constrained and therefore make the assessment of the dust environment challenging.\\\\}
   {\noindent\textbf{Aims:} The main goal of the work is to provide realistic estimations of a dust environment based on very general parameters of possible target objects.\\\\}
   {\noindent\textbf{Methods:} Contemporary numerical models of dusty-gas coma are used to obtain spatial distribution of dust for a given set of parameters. By varying parameters within a range of possible values we obtain an ensemble of possible dust distributions. Then, this ensemble is statistically evaluated in order to define the most probable cases and hence reduce the dispersion. This ensemble can be used to estimate not only the likely dust abundance along e.g. a fly-by trajectory of a spacecraft but also quantify the associated uncertainty.\\\\}
   {\noindent\textbf{Results:} We present a methodology of the dust environment assessment for the case when the target comet is not known beforehand (or when its parameters are known with large uncertainty). We provide an assessment of dust environment for the CI mission. We find that the lack of knowledge of any particular comet results in very large uncertainties ($\sim3$ orders of magnitude) for the dust densities within the coma. The most sensitive parameters affecting the dust densities are the dust size distribution, the dust production rate and coma brightness, often quantified by Af$\rho$. Further, the conversion of a coma's brightness (Af$\rho$) to a dust production rate is poorly constrained. The dust production rate can only be estimated down to an uncertainty of $\sim0.5$ orders of magnitude if the dust size distribution is known in addition to the Af$\rho$. All results are publicly available under \url{https://www.doi.org/10.5281/zenodo.6906815}.\\\\}
  {\noindent\textbf{Conclusions:} To accurately predict the dust environment of a poorly known comet, a statistical approach needs to be taken to properly reflect the uncertainties. This can be done by calculating an ensemble of comae covering all possible combinations within parameter space as shown in this work.}
\end{abstract}

\section{Introduction} \label{sec:introduction}
To date, only seven comets (1P/Halley, 19P/Borrelly, 9P/Tempel 1, 26P/Grigg-Skjellerup [no nucleus imaging], 67P/Churyumov-Gerasimeko, 81P/Wild 2, and 103P/Hartley 2) have been visited by spacecraft.
Spacecraft observations provide a detailed, high spatial and temporal resolution of the surface and surrounding coma but are limited to a few target comets ($\sim\!1-2$ per decade).

While these observations are invaluable to understanding the building blocks of our Solar System and how their activity works exploring them in close proximity bares potential risk for spacecraft.
This is true in particular when a spacecraft passes in close proximity (a few hundred kilometres or closer) of an active comet nucleus during a high velocity fly-by.

When comets enter the inner Solar System and heat up sufficiently, their ices begin to sublimate.
Additionally, the emitted gas drags dust particles into space and they form the so-called gas and dust comae.
Depending on the spatial and temporal scales the dynamics of the dust particles is primarily governed by the interaction with the gas, gravity from the nucleus and the Sun, and solar radiation pressure.
These forces shape the distribution of dust in the vicinity of the comet nucleus as well as far into the tail of a comet millions of kilometres from the comet's surface.

During a fly-by through the coma, dust particles can collide with the spacecraft.
Depending on the impact energy such impacts can pollute the surface and instruments and/or disturb attitude control [e.g. $4$ particles between $0.5$ and $10$ mg impacted the Deep Impact Impactor before it impacted Tempel $1$ \citep{AHearn2008}; $9$ particle ($\sim0.01$ to $\sim0.1$ mg) impacts on the Deep Impact Flyby s/c during its flyby of Hartley $2$ \citep{Hermalyn2013}; and several discrete events caused by dust particles in the $1-50$~mg range were reported by HMC on Giotto \citep{Curdt1990Icar}] up to severe damage to the spacecraft (high impact energy i.e. when the dust mass and/or impact speed are high, e.g. loss of HMC camera on Giotto due to large dust particle impact).
It is crucial to quantify the dust environment to
\begin{itemize}
    \item guide trajectory design,
    \item inform spacecraft shielding, 
    \item assess the performance of star trackers,
    \item evaluate the performance of scientific instruments (like dust sensors/accumulators), and 
    \item determine the stability of attitude control.
\end{itemize}

ESA's Comet Interceptor (CI) mission \cite{SnodgrassJones2019} will pass through a potentially hazardous region of the inner coma of a dynamically new comet or long period comet. 
This is an indispensable condition to get better resolution from remote observations and to collect in-situ data on dust and gas environment.
It is therefore important to assess the dust impact risk to the spacecraft and their scientific instruments to aid hazard mitigation strategies.

This kind of assessment has a long history dating back to e.g. the assessment of the fly-by at comet 1P/Halley by the VEGA spacecraft \citep{Sagdeev1982}.
Other studies have been performed to quantify the dust environment of comet 9P/Tempel 1 for the Deep Impact mission \citep{Lisse2005}, for the Rosetta mission to comet 67P/Churyumov-Gerasimenko \citep{Agarwal2007}, and most recently for potential missions to e.g. active centaurs \citep{Fink2021}.

Though such models describing the dust environment for space missions are not new, the CI mission is in a unique situation.
It is the first mission for which the precise target (a specific comet) of the mission could remain unknown until after launch.
Naturally this is a particular problem for determining the expected dust coma because of the many parameters with a broad range of possible values.
The primary problem is thus not what model to use but what values for the parameters to assume in a model of the inner dust coma.
Furthermore, because the different assumptions about the dust mass loss from a cometary nuclei are strongly interdependent \cite{Marschall2020b} it is not even obvious a priori which set of parameters represent the best/worst case scenarios.

We present here the Engineering Dust Coma Model (EDCM) which has a limited number of input parameters but retains general physical realism of the dust distribution in the inner coma.
The purpose of this model is to make predictions of which dust the three spacecraft of CI will encounter during the active phase of the mission. 

It is important to note, that due to the uncertainty of the target comet and therefore its parameters, we have to use a new approach to building the model.
Instead of defining one set of parameters that shall represent a best, nominal, or worst case scenario we choose ranges for each parameter of the model based on our knowledge of comets as e.g. 1P/Halley and other comets.
All self-consistent combinations within those ranges will be run through our model to give a prediction of all possible coma environments within parameter space.
This ensemble of dust environments are subsequently statistically evaluated to determine a probabilistic distribution of possible conditions which the spacecraft might encounter.

We will describe the general philosophy behind the model and the EDCM in Sec.~\ref{sec:model} and how it was calibrated (Sec.~\ref{sec:calibration}).
Section~\ref{sec:resultsCometInterceptor} will present the results for the spacecraft of the Comet Interceptor mission.
We will also describe how the presented results can be scaled to different fly-by trajectories (Sec.~\ref{sec:scalingResults}).
We end this work by providing a discussion and conclusions (Sec.~\ref{sec:discussion}).

%%%%%%%%%%%%%%%%%%%%%%%%%%%%%%%%%%%%%%%%%%%%%%%%%%%%%%%%%%%%%%%%%%%%%%%%%%
%%%%%%%%%%%%%%%%%%%%%%%%%%%%%%%%%%%%%%%%%%%%%%%%%%%%%%%%%%%%%%%%%%%%%%%%%%
%%%%%%%%%%%%%%%%%%%%%%%%%%%%%%%%%%%%%%%%%%%%%%%%%%%%%%%%%%%%%%%%%%%%%%%%%%
%%%%%%%%%%%%%%%%%%%%%%%%%%%%%%%%%%%%%%%%%%%%%%%%%%%%%%%%%%%%%%%%%%%%%%%%%%
%%%%%%%%%%%%%%%%%%%%%%%%%%%%%%%%%%%%%%%%%%%%%%%%%%%%%%%%%%%%%%%%%%%%%%%%%%
%%%%%%%%%%%%%%%%%%%%%%%%%%%%%%%%%%%%%%%%%%%%%%%%%%%%%%%%%%%%%%%%%%%%%%%%%%
\section{The model}\label{sec:model}
\subsection{A shift in the modelling approach}
For the assessment of the dust coma environment, missions to particular comets (e.g. previous missions to comets 1P/Halley, or 67P/Churyumov-Gerasimenko) had a significant advantage over CI.
Their targets were known before launch and ground-based data with information about the targets could inform the expected dust coma environment \citep{Sagdeev1982,Lisse2005,Agarwal2007,Fink2021}.

In contrast the target of CI will be selected shortly before or even after the spacecraft have been launched into space \citep{SnodgrassJones2019}.
Therefore, it is necessary to assess the dust environment for a range of conditions which CI might encounter. 

Two issues are of particular importance in our case.
First, because the target is unknown, or at the least it's activity within the water ice line and in particular at the encounter distance of 1~au, so are the properties about its activity.
We don't know how large the nucleus will be, what the gas and dust production rates are around 1 au, or what the dust size distribution is.
Second, more fundamentally, even if some of these properties were known the different dust properties are inter-dependant in a way that makes it unclear a priori which set of parameters represent a best, nominal, or worst case scenario.

We illustrate this with a simple example of just three parameters influencing the dust environment: the gas production rate, the dust-to-gas mass flux ratio, and the dust size distribution.
Each of these individually trivially affects the dust densities within the coma.
For a given dust-to-gas mass flux ratio and dust size distribution, an increase in the gas production rate will result in an increase of the dust densities in the coma.
Likewise, for a given gas production rate and dust size distribution, an increase in the dust-to-gas mass flux ratio will increase the dust densities in the coma.
Finally, the abundance of a specific size can be directly determined by the dust size distribution provided a given gas production rate and dust-to-gas mass flux ratio. 

While these relationships are intuitive, combinations of them no longer have an obvious result.
For example which of the following two combinations has more large particles?
\begin{enumerate}
    \item a coma with a high gas production rate, low dust-to-gas mass flux ratio, and size distribution dominated by large particles,
    \item a coma with a high gas production rate, high dust-to-gas mass flux ratio, and size distribution dominated by small particles.
\end{enumerate}
Of course this depends on the exact values of each of these three parameters but it illustrates the fundamental dilemma.
And this is not the full extent of the problem.
In reality we are not confronted by only three inter-dependant parameters but as we will discuss below by at least 11.

It is therefore unclear how to properly define a set of parameters which shall correspond to a certain scenario (best, nominal, worst).
For this reason, we have decided to present a new approach to this problem.
While previous work could rely on a narrow set of input parameters we turn this around.

For each parameter of the model we will consider a broad range of values which are compatible with known observations of comets.
Within this parameter space we select all possible combinations that are self-consistent. 
For example a parameter combination can be excluded if it were to result in an Af$\rho$ higher than the upper limit assumed here.
This set of self-consistent parameter combinations will form the basis of the model.

Each self-consistent parameter combination is then run through our dust coma model and results in a unique coma environment.
The resulting set of coma environments can then be analysed in a statistical manner to determine e.g. a nominal case.

Our approach thus turns around the process of predicting the coma environment.
Instead of defining a nominal case, we run all plausible parameter combinations and let the model tell us which dust densities are expected to be encountered.

\subsection{Overview of the model components}
\begin{figure*}[h]
	\includegraphics[width=\textwidth]{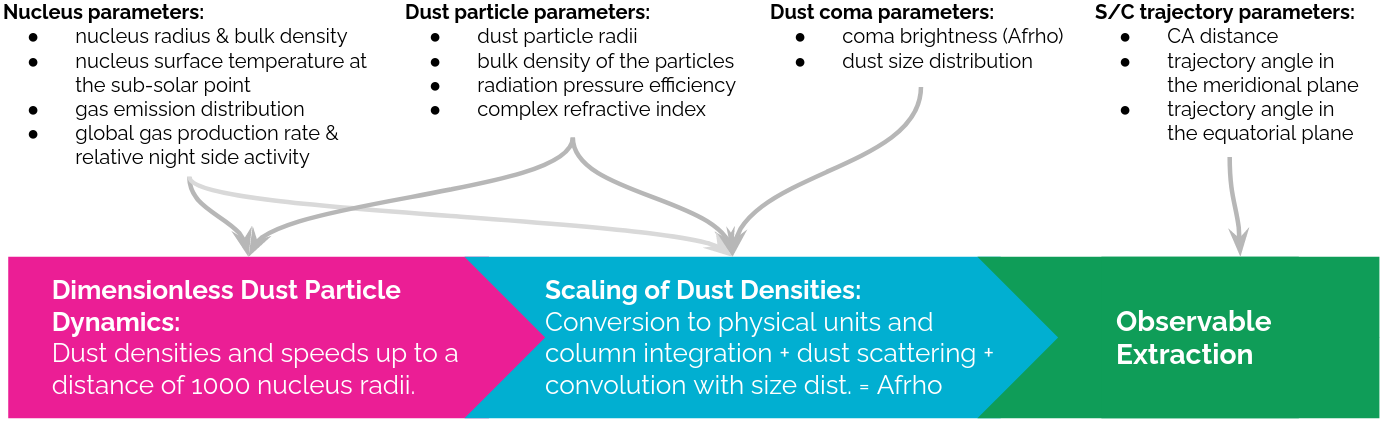}
	\caption{Sketch of the components of the EDCM and which input parameters feed into which step.}
	\label{fig:sketch} 
\end{figure*}

\noindent The EDCM is composed of three components: 
\begin{itemize}
    \item the {\it{dimensionless dust particle dynamics}} which calculates the spatial distribution of dust,
    \item the {\it{scaling of dust volume densities}} which makes the conversion to physical units and performs column integration taking into account dust scattering and a size distribution in order to obtain the brightness Af$\rho$ \citep[defined by ][it is the product of grain albedo ($A$), the cross sectional filling factor of the dust within the photometric aperture ($f$), and the circular aperture's radius ($\rho$).  For idealized circumstances (constant production, radial outflow) it is proportional to the dust mass loss rate \citep{FinkRubin2012}.]{AHearn1984AJ} and make the selection of self-consistent cases,
    \item the {\it{observable extraction}} that determines the number density encountered along a specific spacecraft trajectory and performs the estimation of the probability of encountering the respective case.
\end{itemize} 
The three components are depicted in magenta, blue, and green, respectively in Fig.~\ref{fig:sketch} and will be described in detail in the following sections.
Which parameters are required by which component are illustrated as well.

Table~\ref{tab:parameters} lists the parameters, respective symbols, and ranges used for the target of the CI mission.
The ranges used have been chosen such that CI will be able to survive a fly-by similar to the one by Giotto at comet 1P/Halley (see discussion later on).

Some parameters are of note.
For now we have fixed the heliocentric distance to $1$~au (where the CI fly-by will occur) thus also fixing the nucleus surface temperature at the sub-solar point.
The model will be valid for small variations in the heliocentric distance ($\pm20\%$).
This is because of the large range of possible gas production rates we are allowing in the model.
To some extent a lower heliocentric distance resulting in a higher production rate is therefore indirectly contained in the model.
The nucleus is assumed to be a homogeneous sphere of radius $R_N$.
We assume 2 and 5~km to cover the range of comets like 67P \citep[][]{Preusker2017} to that of comet Halley \citep[][]{Keller1987,Wilhelm1987}.
The bulk nucleus density was assumed as low as for objects like the cold classical Arrokoth \citep[][]{Spencer2020} up to the canonical value of 500~kg~m$^{-3}$ for comets \citep[][]{Preusker2017}

The gas emission distribution, $q_g(\varphi)$, is assumed to be proportional to $cos^k(\varphi)$, where $\varphi$ is the solar zenith angle, and the power $k$ takes the values 1, 2, and 3 (which corresponds to the different mechanisms of gas production from the nucleus).
For example \citet{Zakharov2021} showed that at 1~au sublimation from the surface results in an emission distribution closely following the case where $k=1$ \citep[Fig. 2 in][]{Zakharov2021}.
If on the other hand, sublimation occurs inside the surface layer then the distribution changes to the case where $k=3$.
In this sense, $k$ can be interpreted as the parametrisation of where sublimation occurs.

The gas production rate is assumed for an equivalent sphere with a radius of $1$~km and then scaled according to the actual nucleus radius used.
The lower limit corresponds to a comet with an activity equivalent to comet 67P at 1.4~au inbound \citep[][]{Combi2020,Marschall2020b} and the upper limit to roughly twice that of comet 1P/Halley during the Giotto fly-by \citep[][]{Krankowsky1986}.
Night-side activity is introduced as uniform emission with respect to $\varphi$ and quantified by the ratio of the gas production rate at the anti-solar point to the production rate at the sub-solar point, $q_g^{180^{\circ}}/q_g^{0^{\circ}}$.

We allow for a large range of dust particles from micron to decimetre sized which have been observed at comets \citep[][]{McDonnell1997,Price2010,Mannel2016,Merouane2017,Ott2017,Fulle2016,LevasseurRegourd2018SSRv,DellaCorte2019,Mannel2019}.
The slope of the size distribution has also been chosen according to previous measurements \citep[e.g.][]{McDonnell1997,Price2010,Merouane2017}.
The brightness of the coma, expressed in Af$\rho$ was chosen between 2,800~cm and 28,000~cm with the upper limit being slightly higher than that of comet 1P/Halley during the Giotto encounter \citep[][]{Fink1994,Schleicher1998}.

\begin{table*}[h]
\caption{List of parameters used in the EDCM for the CI mission target.} \label{tab:parameters}
\begin{tabular}{p{0.1\textwidth}p{0.07\textwidth}|p{0.43\textwidth}|p{0.2\textwidth}|p{0.1\textwidth}}
symbol      & unit      & parameter                                         & value/ range/ function  & step size\\\hline\hline
&&\textbf{Nucleus parameters}&\\\hline
$r_h$       & au        & heliocentric distance                             &  1     & n/a\\
$R_N$       & km        & nucleus radius                                    &  {2, 5}  & n/a   \\
$\rho_N$    & kg m$^{-3}$ & bulk density of the nucleus                     &  {317, 508} & n/a \\
$T_N$       & K         & nucleus surface temperature at the sub-solar point&  317   & n/a  \\
$q(\varphi)$& -         & gas emission distribution ($\sim cos^k(\varphi)$)&   $k=1,2,3$ & n/a    \\
$Q_g^{equiv}$& s$^{-1}$ & global gas production rate for an equivalent sphere with $R=1$~km ($Q_g^{equiv} = Q_g /R_N^2$)&  \textbf{$1.3\cdot10^{27}$} $- 4.9\cdot10^{28}$ & uniformly sampled     \\
$q_g^{180^{\circ}}/q_g^{0^{\circ}}$ & - & relative night side activity            & 0.01, 0.05, 0.1 & n/a\\\hline
&&\textbf{Dust particle parameters}&\\\hline
$a$         & m         & radius                              & $10^{-6} - 10^{-1}$ & half decade    \\
$\varrho_d(a)$ & kg m$^{-3}$ & bulk density           & $300-1000$ & $100$      \\
$q_{rad}(a)$& -         & radiation pressure efficiency                     & $0.2-2$ & $\sim 0.1$     \\
$q_{sca}(a)$& -        & scattering efficiency                             &  silicates \& organics$^*$  & n/a    \\
$\phi(\alpha)$& -      & phase function                               &  silicates \& organics$^*$  & n/a     \\\hline
&&\textbf{Dust coma parameters}&\\\hline
Af$\rho$   & cm        & brightness                       &  {$2,800 - 28,000$} & $2,000$ \\
$\beta$    & -         & differential power law exponent of the dust size distribution; $n \sim a^{-\beta}$  &  $3.2-4.4$ & 0.2 \\
$\chi$*     & -         & dust-to-gas mass production rate ratio            &  $< 10$   & n/a \\\hline

&&\textbf{S/C trajectory parameters}&\\\hline
$r_{CA}$   & km        & cometocentric distance of spacecraft at closest aThe closest approach (CA) distance, the trajectory angle in the meridional plane, $\alpha_T$, and the trajectory angle in the equatorial plane, $\beta_T$.

pproach (CA)                       &  1000, 500, 200 & for S/C A, B1, B2 \\
$\alpha_{T}$    & deg        & trajectory angle in the meridional plane &  $0$ & n/a \\
$\beta_{T}$    & deg       & trajectory angle in the equatorial plane  &  $0$ & n/a \\\hline 

\end{tabular}
\flushleft
Note: $^*$ See \cite{Marschall2020b}.
\end{table*}

\subsection{The dust particle dynamics}
We assume that the dusty-gas coma is formed by the gas sublimating from the nucleus (from the surface and/or from the interior), and solid particles (mineral or/and icy) released from the nucleus with zero initial velocity and entrained by the gas flow.
It is assumed that the dusty-gas flow is coupled in one way only -- the gas drags the dust (i.e. the presence of dust in the coma does not affect the gas motion), and that the dust particles do not collide with each other (the assumption of no feedback on the gas for most comae was articulated in \citet{Tenishev2011}).
The dust particles are assumed to be isothermal, spherical and internally homogeneous with invariable size and mass.
To compute the dust distributions we adopt  a frame attached to the nucleus but not rotating with it.
Under these assumptions, dust particle motion is governed by the equations \citep{Crifo2005}:

\begin{equation}
m_d \frac{d\vec v_d}{dt} = \vec F_A + \vec F_G + \vec F_S + \vec F_I
\label{eqn_dvdt}
\end{equation}
\begin{equation}
\frac{d\vec r}{dt}= \vec v_d
\label{eqn_drdt}
\end{equation}
where $t$ is the time, $\vec r$, $\vec v_d$ are the particle radius and velocity vectors, and $m_d$ particle mass, $\vec F_A$ is the aerodynamic force, $\vec F_G$ is the nucleus gravity force, $\vec F_S$ is the solar force, and $\vec F_I$ is the inertia force.

For isothermal, spherical and internally homogeneous non-sublimating grains, the aerodynamic force degenerates to a drag force:
\begin{equation}
\vec F_A = \frac{(\vec v_g- \vec v_d)^2}{2} \frac{(\vec v_g- \vec v_d)}{|\vec v_g- \vec v_d|} \rho_g \sigma_d C_D\left(s,\frac{T_d}{T_g}\right)
\label{eqn_fa}
\end{equation}
where $\sigma_d$ and $T_d$ are the particle cross-section and temperature, $\rho_g$, $v_g$, $T_g$ are the gas density, velocity and temperature, $C_D$ is the drag coefficient. 
For a spherical particle $\sigma_d=\pi a^2$ and $m_d=\frac{4}{3}\pi a^3 \varrho_d$, where $a$ is the particle radius and $\varrho_d$ is the particle specific density.
If the particle size $a$ is much smaller than the mean free path of the gas molecules (typical case in the coma), the drag coefficient $C_D$ can be estimated by the ``free-molecular'' expression for a spherical particle \citep[see][]{Bird1994}:

\begin{align}
C_D \left(s,\frac{T_d}{T_g} \right) &= \frac{2s^2+1}{s^3 \sqrt{\pi}} \exp(-s^2) \nonumber\\
&+\frac{4s^4+4s^2-1}{2s^4} {\rm erf}(s)\\
&+\frac{2 \sqrt{\pi}}{3s}\sqrt{\frac{T_d}{T_g}} \nonumber
\label{eqn_cd}
\end{align}
where $s=\vert \vec v_g- \vec v_d \vert / \sqrt{2 T_g k_B/m_g}$ is a speed ratio, $m_g$ is the mass of the molecule, $k_B$ is the Boltzmann constant.

The nucleus gravitational force $\vec F_G$ is:
\begin{equation}
\vec F_G = m_d \vec g
\label{eqn_fg}
\end{equation}
where $\vec g$ is the gravitational acceleration in the field of the nucleus.
For the spherical nucleus $\vec g =-G M_N\vec r/|r^3|$ where $G$ is the gravitational constant, $M_N$ is the mass of the nucleus.

The solar force $\vec F_S$ which includes solar gravity and solar radiation pressure and is given, for spherical grains, by:
\begin{equation}
\vec F_S = \left(\frac{G M_\odot m_d}{r^2_{h,d} r^2_\oplus} - \sigma_d \frac{Q_{ef} c_\odot \xi(\vec r)}{c r^2_{h,d}}\right) \vec i_{\odot,d}
\label{eqn_fs}
\end{equation}
where $M_\odot$ is the mass of the Sun, $Q_{ef}$ is the radiation pressure efficiency, $c$ is the speed of light, $c_\odot$ is the solar energy flux at 1AU, $r_h$ is the heliocentric distance of the nucleus (in AU),
$r_{h,d}$ is the dust particle heliocentric distance (in AU),
$r_\oplus = 1.496\cdot10^{11}$~m is the Earth orbit semimajor axis,
$\vec i_{\odot,d}$ is the unit vector directed from the dust particle to the Sun, and $\xi (\vec r)$=0 if $\vec r$ is inside the shadow, and $\xi (\vec r)$=1 if $\vec r$ is in sunlight.

In the adopted nucleus-attached (but non rotating) frame, the inertia force $\vec F_I$ is given \citep{Landau1976} by:
\begin{equation}
\vec F_I = m_d \frac{G M_\odot}{r^2_h r^2_\oplus} \vec i_{\odot,n} 
\label{eqn_fi}
\end{equation}
where $\vec i_{\odot,n}$ is the unit vector from the nucleus center-of-mass to the Sun.

Following \citet{Zakharov2018b} and \citet{Zakharov2021b} we introduce the following dimensionless variables:
$\theta=T_d/T_s$, $\tilde v_g = v_g/v_g^{max}$, $\tilde v_d = v_d/v_g^{max}$, $\tilde r = r/R_N$, $\tilde t = t/\Delta t$, $\tau=T_d/T_g$, $\tilde \rho_g=\rho_g / \rho_s$.
Here $v_g^{max}=\sqrt{\gamma \frac{\gamma+1}{\gamma-1} \frac{k_b}{m_g} T_s}$ is the theoretical maximum velocity of gas expansion, $\rho_s$ and $T_s$ are the gas density and temperature on the sonic surface (i.e. where the gas velocity is equal the local sound velocity $\sqrt{\gamma k_B T_g/m_g}$) at the sub-solar point, $\gamma$ is the specific heat ratio, $R_N$ is the characteristic linear scale (e.g. the radius of the nucleus), and $\Delta t=R_N/v_g^{max}$. 
In addition, we introduce $\tilde g$ via normalizing $g$ on $G M_N/R_N^2$ (therefore, in the case of a spherical nucleus $\tilde g =-\vec {\tilde r}/|{\tilde r}^3|$).

Using these dimensionless variables, the equations (\ref{eqn_dvdt}) and (\ref{eqn_drdt}) can be rewritten as:

\begin{align}
\frac{d\vec{\tilde v}_d}{d\tilde t} = & \tilde \rho_g (\vec{\tilde v}_g - \vec{\tilde v}_d)^2 \frac{\vec{\tilde v}_g - \vec{\tilde v}_d}{|\vec{\tilde v}_g - \vec{\tilde v}_d|} C_D(s,\tau) ~ {\rm Iv} + {\rm Fu} ~ \vec{\tilde g} + \nonumber\\
& \frac{R_N G M_\odot}{(v_g^{max})^2 r^2_h r^2_\oplus}\times \nonumber\\ 
&\left[\left(1-\frac{3 c_\odot r^2_\oplus}{4 G M_\odot c} \frac{Q_{ef}}{\varrho_d a} \xi(\vec{\tilde r})\right)\frac{r^2_h}{r^2_{h,d}} \vec i_{\odot,d} -\vec i_{\odot,n}\right] \label{eqn_dvtdtt0}
\end{align}

\begin{equation}
\frac{d\vec{\tilde r}}{d\tilde t}= \vec{\tilde v}_d \qquad{}\qquad{}\qquad{}\qquad{}\qquad{}\qquad{}\qquad{}\qquad{}\qquad{}\qquad{}
\label{eqn_drtdtt}
\end{equation}

If the linear size of computational domain is much smaller than $r_h$, then $r_h \simeq r_{h,d}$ and $\vec i_{\odot,d} \simeq \vec i_{\odot,n}$. In this case the equation \ref{eqn_dvtdtt0} may be rewritten as:

\begin{equation}
\frac{d\vec{\tilde v}_d}{d\tilde t} = \tilde \rho_g (\vec{\tilde v}_g - \vec{\tilde v}_d)^2 \frac{\vec{\tilde v}_g - \vec{\tilde v}_d}{|\vec{\tilde v}_g - \vec{\tilde v}_d|} C_D(s,\tau) ~ {\rm Iv} + {\rm Fu} ~ \vec{\tilde g} - {\rm Ro} ~ \xi(\vec{\tilde r}) \vec i_{\odot,n}
\label{eqn_dvtdtt}
.\end{equation}
As a consequence of this assumption, this paper neglects the solar tidal forces that affect significantly the dust motion out of the acceleration zone over long times (weeks and longer).
For particles equal and larger than mm-sized these effects are important also just outside the dust acceleration zone.
The rigorous computation in the heliocentric reference frame of the dust motion out of cometary Hill's sphere can be found in \citet{Fulle1995}.

The equations \ref{eqn_dvtdtt0} and \ref{eqn_dvtdtt} contain three dimensionless parameters:
\begin{equation}
{\rm Iv} = \frac{1}{2} \frac{\rho_s \sigma_d R_N}{m_d} = \frac{3}{8} \frac{\rho_s R_N}{a \varrho_d} = \frac{3 Q_g^{sph} m_g}{32 R_N a \varrho_d \pi \sqrt{T_s \gamma k_B/m_g}}
\end{equation}
and
\begin{equation}
{\rm Fu} = \frac{G M_N}{R_N} \frac{1}{(v_g^{max})^2}
\end{equation}
and
\begin{equation}
{\rm Ro} = \frac{1}{m_d (v_g^{max})^2} R_N \frac{\sigma_d Q_{ef} c_\odot}{c r_h^2}
.\end{equation}
Here $Q_g^{sph}$ is the total gas production rate [s$^{-1}$] of a spherical nucleus of radius $R_N$ with the local gas production uniform over the surface and equal the gas production in sub-solar point.

We note that the theoretical maximum velocity of gas expansion could be expressed also in terms of a stagnation temperature $T_0$ and a heat capacity $C_p$ as $v_g^{max}=\sqrt{2 C_p T_0}$. 
In this way the gas velocity on the sonic surface is $v_s= \sqrt{2 C_p T_0} \sqrt{\frac{\gamma-1}{\gamma+1}}$ and the local gas production is $q = \rho_s v_s$. 
Therefore we can rewrite $\rm Iv$ as:
\begin{equation}
{\rm Iv} = \sqrt{\frac{\gamma+1}{2(\gamma-1)}} \frac{3 q R_N}{8 \sqrt{C_p T_0} a \varrho_d}
\label{eqn_Iv2}
.\end{equation}

The second term in this equation is the reciprocal of a dimensionless similarity parameter characterising the ability of a dust particle to adjust to the local gas velocity introduced in \citet{Probstein1969}.

These parameters have the following meaning:
\begin{enumerate}
\item $\rm Iv$ represents the ratio of the gas mass present in a flow tube, with the cross section of the particle and a characteristic length $R_N$, to the particle mass. 
This parameter characterises the efficiency of entrainment of the particle within the gas flow (i.e. the ability of a dust particle to adjust to the gas velocity);
\item $\rm Fu$ represents the ratio of the comet surface gravitational potential to the flow thermodynamic potential (enthalpy, $C_p T_0$).
This parameter characterises the efficiency of gravitational attraction;
\item $\rm Ro$ represents the ratio of the specific work done by the solar pressure force on the characteristic length $R_N$ to the flow thermodynamic potential. This parameter characterises the contribution of solar radiation pressure.
\end{enumerate}
In order to define $\rm Iv, Fu, Ro$ it is necessary to know: $m_g$, $\gamma$, $\rho_s$, $T_s$, $R_N$, $M_N$, $a$ and $\varrho_d$ (or $\sigma_d$ and $m_d$), $Q_{ef}$ and $r_h$.

We have computed the dust distribution within the cometary coma up to 1000 nucleus radii ($R_N$).
It uses a minimal number of parameters for the description of a cometary dust coma, while keeping it physically realistic.
This model physically consistently takes into account the expanding nature and asymmetry of the gas coma (caused by gas production modulated by solar radiation) and considers the dust dynamics driven by the gas drag force, force from the nucleus gravity, and solar radiation pressure. 

A series of general assumptions were made to simplify the model. 
For each of the simplifications we also outline the resulting limitations that need to be appreciated.
\begin{itemize}
    \item \textbf{The nucleus shape is assumed to be spherical.} 
    The dust dynamical model can therefore not reproduce possible heterogeneity within the coma that arise due to topography (and often occurs only at particular orientation of the nucleus with respect to the Sun). 
    Nevertheless, when the parameters of the coma of a complex shape nucleus are averaged over a full comet rotation the resulting average coma distribution resembles closely the one from a spherical nucleus \citep[e.g.][]{Zakharov2021}.
    \item \textbf{The gas is assumed to be an ideal perfect gas.} 
    The gas flow in the coma is described by the gas-dynamic Euler equations which express the conservation of the mass, momentum and energy in the flow of an ideal perfect gas.\\
    In other words these equations describe the motion of an equilibrium fluid flow without viscous dissipation and heat conductivity. The real atmosphere of a comet contains rarefied non-equilibrium regions as well. \\
    The transfer of thermal energy into kinetic energy in a rarefied flow is less efficient than in a fluid flow, therefore the rarefied flow accelerates slower. Due to the presence of viscous dissipation flow structures like shocks are diffused in a rarefied flow.\\
    Nevertheless, as was shown in many publications, the description of the flow from a spherical nucleus based on the fluid equations preserves general physical realism of the flow.
    As was noticed in \citet{Crifo2004}, the existing results from comparisons of kinetic and fluid approaches (e.g. \citet{Crifo2002a} and \citet{Crifo2003}) show that the Navier-Stokes equations and even the Euler equations provide acceptable solutions over practically the whole day-side coma of observable comets.
    Two restrictions are to be made, however: (1) the immediate vicinity of the nucleus surface (several $R_N$), and (2) the outer reaches of the coma, where dissociation products are dominant, it is presently not known to which accuracy these equations represent the real situation. \\
    An important merit of the Euler equations is that they don't depend on the flow rarefaction and therefore the solutions can be scaled for different production rates, the main reason why we use them in the present model. 
    \item \textbf{It is assumed that the dust does not influence the gas flow} (i.e. no back-coupling of the dust to the gas flow). 
    This allows the separate/sequential determination of the gas and dust flows.
    This is given for comets with low dust-to-gas ratios [$\sim < 10$]. Nevertheless, for a comet with a high dust-to-gas ratio the model will be able to produce reliable predictions.
    \item \textbf{It is assumed that the gas coma is constituted of one single species, H$_2$O.} 
    This assumption will be reasonable well satisfied for comets where H$_2$O is the dominant gas species (as e.g. for comet 67P, most comets at 1~au). 
    \item \textbf{It is assumed that there is no extended gas/dust source/sink in the coma.} 
    The model should not be applied to comets with a significant extended gas/dust source/sink e.g. emission of gas from dust particles within the coma, sublimation and destruction of dust particles etc.
    \item \textbf{It is assumed that the gas and dust emission is smooth across the surface or that any inhomogeneities are blurred within the drag acceleration region $r \leq 10R_N$.} 
    The gas and dust emission is modulated e.g. by solar zenith angle. 
    This means that the emission \textbf{is not} dominated by a few very localised jets. 
    \item \textbf{The model is reliable for cometo-centric distances of $10R_N<r<1000R_N$.} 
    The inner boundary is defined by the possible existence of 'fine structures' of the flow due to particularities of the surface topography. 
    The outer boundary is defined by the size of the computational region. 
    It is possible to extrapolate the data beyond this upper limit via introducing additional assumptions (e.g. sphericity of expansion etc.).  
    \item \textbf{The model is run for a heliocentric distance of $1$~AU}, the approximate distance of CI's comet during the encounter. 
    Result shall thus not be attempted to be scaled to vastly different heliocentric distances. 
    The results remain valid for variations of the heliocentric distance that do not change the solar flux to the nucleus surface by more than a factor of 2 (i.e. the model is valid for the heliocentric distance range of CI).
    \item \textbf{It is assumed that dust particles are spherical, homogeneous with invariable size and mass.}
\end{itemize}

This simplification allows us to solve the equation of motion generally in (Iv, Ro, Fu) space and then re-scale these results to the respective physical values in a later step.
For all details on this model we refer the reader to \cite{Zakharov2018b} and \cite{Zakharov2021b}.

For the underlying gas dynamics model we used the results by \cite{Zakharov2021} who have calculated the gas field by solving the Euler equations. 
The gas results are used to calculate the dynamics of spherical dust particles taking into account gas drag, nucleus gravity, and solar radiation pressure \citep[Eq.~\ref{eqn_dvdt};][]{Zakharov2018b,Zakharov2021b}.
An important implicit assumptions to reiterate is that we assume that the dust does not have a back reaction effect on the gas flow.

The dust dynamics makes use of dimensionless variables (derived by combining the respective physical parameters from Table \ref{tab:parameters}) to parametrise the dust dynamics as described in \cite{Zakharov2018} and \cite{Zakharov2021b} and reduce the number of numerical solutions. 
The result of this step is the dust number density and velocity in 3D space.

At this point there is also the implicit assumption that the dust-to-gas mass production rate ratio, $\chi$, is unity.
These solutions therefore do not yet have an absolute scale and thus requring the second component of the EDCM.\\

\subsection{Scaling the model to physical parameters}
In a first step we convert the dimensionless results given in (Iv, Ro, Fu)-space into physical units.
This is done for each of the parameter combination in our parameter set.

Once this is done only one parameter has yet to be fixed.
Until now the dust-to-gas ratio, $\chi$, is still assumed to be one.
$\chi$ will be chosen such that a certain brightness of the coma, Af$\rho$, is reached.
The larger $\chi$ the larger Af$\rho$.
But we need to identify what value of $\chi$ results in which Af$\rho$.
The calculation of Af$\rho$ for each parameter combination follows the approach described in \cite{Marschall2016}. 

First, the dust column density of an aperture of $20R_N$ is calculated. 
For points outside the simulation domain ($> 1000 R_N$) a $1/r^2$ extrapolation is applied.
The column densities are then weighted with a power law ($n \sim a^{-\beta}$) and converted into reflectance using the scattering model of \cite{Markkanen2018} as shown in \cite{Marschall2020b}.

Second, the resulting reflectance can then be used to calculate the Af$\rho$ as explained in \cite{Gerig2018}.
The absolute scaling $\chi$ can then be determined by linearly scaling the model Af$\rho$ to the desired Af$\rho$. I.e. if the model Af$\rho=100$~cm then an actual coma with Af$\rho=200$~cm is achieved with $\chi=2$.

Two additional constraints are added at this point:
\begin{enumerate}
    \item $\chi<10$, and
    \item $-29.7<$log(Af$\rho$/Q$_{H_2O}) <-27.65$.
\end{enumerate}

The second constraint comes from ground-based observations of comets \citep{AHearn1995,SchleicherBair2016DPS}.
These measurements show that comets don't exhibit an arbitrary range of Af$\rho$/Q$_{H_2O}$.

\begin{figure}[h]
	\includegraphics[width=\columnwidth]{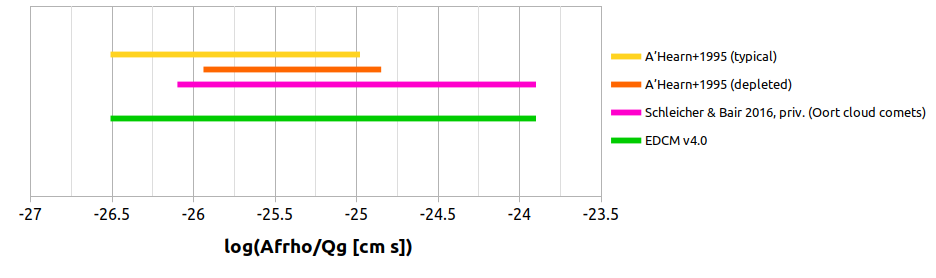}
	\caption{Shows the log(Af$\rho$/Q$_{H_2O}$) from \cite{AHearn1995} for typical (yellow) and depleted (orange) comets as well as the range of the EDCM.}
	\label{fig:afrhoQ} 
\end{figure}

Figure~\ref{fig:afrhoQ} shows the ranges for typical (yellow) and depleted (orange) comets according to Table~VI in \cite{AHearn1995}. 
The range covered in this work corresponds to a range spanning the ones from \cite{AHearn1995,SchleicherBair2016DPS}, and Schleicher \& Bair (private communication; Oort cloud comets).

\begin{figure}[h]
	\includegraphics[width=\columnwidth]{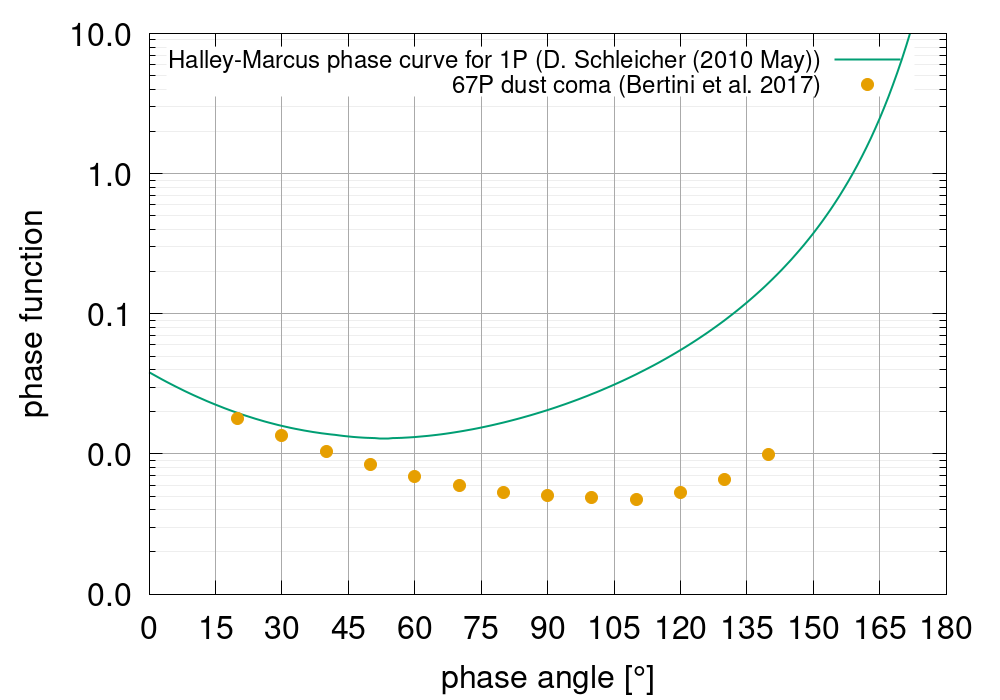}
	\caption{The phase curves of 1P (Schleicher) and 67P \citep{Bertini2017} are shown.}
	\label{fig:phaseFunction} 
\end{figure}

We have chosen to calibrate the model on the Giotto data (see Sec.~\ref{sec:calibration}) for the comparison but employed a scattering model calibrated to comet 67P \citep[][]{Markkanen2018}. 
This introduces some bias because the phase functions of 1P and 67P differ significantly, especially at large phase angles (Fig.~\ref{fig:phaseFunction}). 
Note that these two phase curves have been measured differently (1P from ground, 67P from inside the coma).
The Halley-Marcus phase curve (or Schleicher-Marcus phase curve) is a composite.
The part originating from Halley only extended up to ~55$^{\circ}$ \citep[][]{Schleicher1998}, and the remainder is based on a Henyey-Greenstein function that is consistent with some near-Sun comets \citep[][]{Marcus2007}\footnote{See \url{https://asteroid.lowell.edu/comet/dustphase_details.html}}.

Because we employ the scattering model by \cite{Markkanen2018} for the calculation of the model Af$\rho$, a global scaling factor of $23$ is applied to bring in line the Af$\rho$ measurements of 1P with the Giotto dust densities. 
Therefore, by definition our model fits the Halley data when the appropriate input parameters for Halley are chosen. 
Thus all results can be understood as relative to Halley. 
In this sense if the model is applied to e.g. a comet that is otherwise identical to Halley but has a dust-to-gas ratio that is double that of Halley, the model will predict densities that are double those measured with Giotto.

\subsection{Extracting observables}
Now that our set of dust comae have been scaled to the respective physical units we are left with the final component of the EDCM.
We can now extract the number density encountered along a specific spacecraft trajectories for coma in our set of input parameters.
Again, for points outside the simulation domain ($>1000 R_N$) a $1/r^2$ extrapolation is applied.

\begin{figure}[h]
	\includegraphics[width=\columnwidth]{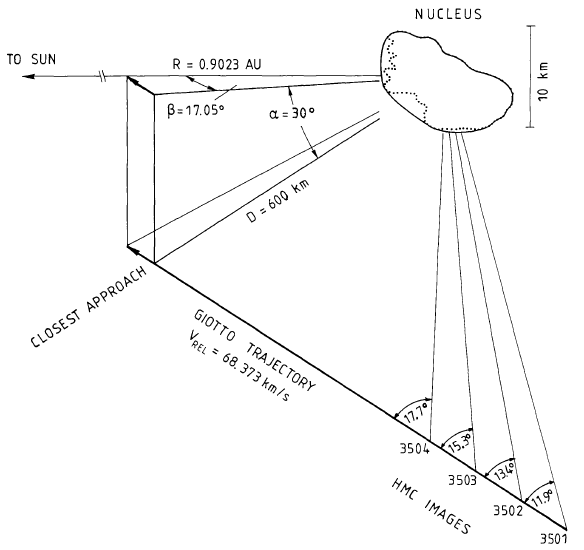}
	\caption{The Giotto flyby trajectory relative to the cometary nucleus and Sun is shown, including offset angles and positions for the last 4 useful HMC images (from \cite{Keller1987}).}
	\label{fig:giotto_trj} 
\end{figure}

The trajectories are defined by three parameters (Table~\ref{tab:parameters}).
The closest approach (CA) distance, the trajectory angle in the meridional plane, $\alpha_T$, and the trajectory angle in the equatorial plane, $\beta_T$.
The nominal CA distances for the three CI spacecraft A, B1, and B2 are 1000, 500, and 200 km respectively.
The two angles are depicted as $\alpha$ and $\beta$ in Fig.~\ref{fig:giotto_trj}.

For our nominal cases we have assumed that the fly-bys of all three spacecraft occur in the same plane, cross the Sun-comet line at $90^{\circ}$, and thus go through $0^{\circ}$ phase angle.
Though these will not be the actual trajectories of the three spacecraft our assumed trajectories are the more conservative assumption.
Trajectories out of the assumed plane result in smaller dust densities (see Sec.~\ref{sec:scalingResults}).
As soon as the trajectories are known the model can be adjusted to the actual trajectory.

We have extracted the dust number density along the trajectory as well as the integrated number of particles along the trajectory (i.e. the column density) for each dust size separately.

In a final step we analyse the results from the set of dust comae.
In particular we have calculated the median column density for each dust size as well as certain percentiles (specifically $5\%$, $10\%$, $25\%$, $75\%$, $90\%$, $95\%$).
This gives a measure for the scatter of cases and thus a measure for the uncertainty of possible outcomes.

%%%%%%%%%%%%%%%%%%%%%%%%%%%%%%%%%%%%%%%%%%%%%%%%%%%%%%%%%%%%%%%%%%%%%%%%%%
%%%%%%%%%%%%%%%%%%%%%%%%%%%%%%%%%%%%%%%%%%%%%%%%%%%%%%%%%%%%%%%%%%%%%%%%%%
%%%%%%%%%%%%%%%%%%%%%%%%%%%%%%%%%%%%%%%%%%%%%%%%%%%%%%%%%%%%%%%%%%%%%%%%%%
%%%%%%%%%%%%%%%%%%%%%%%%%%%%%%%%%%%%%%%%%%%%%%%%%%%%%%%%%%%%%%%%%%%%%%%%%%
%%%%%%%%%%%%%%%%%%%%%%%%%%%%%%%%%%%%%%%%%%%%%%%%%%%%%%%%%%%%%%%%%%%%%%%%%%
%%%%%%%%%%%%%%%%%%%%%%%%%%%%%%%%%%%%%%%%%%%%%%%%%%%%%%%%%%%%%%%%%%%%%%%%%%
\section{Model calibration}\label{sec:calibration}
To calibrate the EDCM we have used the case of comet 1P/Halley which was visited a.o. by the Giotto mission.

The 3-dimensional shape of the comet Halley's nucleus was derived from the combination of Vega and Giotto images. 
A self-consistent model has been produced using an ellipsoid with 8.0, 8.2, and 16 km for the three axes \cite{Wilhelm1987}. 
The surface is about 400 km$^2$ and volume 580 km$^3$. 
The spin vector points about 60$^\circ$ south of the comet to sun direction and it is about perpendicular to the long axis of the nucleus. 
A spin period of 54$\pm$1 h matches the data from the three spacecraft.

In situ measurements \cite{Krankowsky1986} showed a total gas production rate of 

6.9 $\times$ 10$^{29}$ molecules s$^{-1}$ with the predominance of water vapour $\sim$80\% (and 20\% more volatile compounds) by volume and the photodestruction scale length for H$_2$O of 3.9 $\times$ 10$^4$ km. 
\cite{Keller1987} noted that the Halley Multicolour Camera (HMC) images have revealed that the activity of comet Halley's nucleus was concentrated towards the sun and that the variability of the comet during Multi Detector Mode was seen to be small (the minimum-maximum variation is considerably less than a factor of 2).

The results of analysis in \cite{Rickman1989} provide the mass of the nucleus 1.3$\pm$0.3 -- 3.1$\pm$0.4 $\times$10$^{14}$ kg, the mean density of the nucleus 280$\pm$100 -- 650$\pm$190 kg/m$^3$, and the dust-to-gas ratio 0.3--1.

The EDCM uses axially-symmetric numerical solutions. 
Therefore, some trajectories which are different in 3D space might be identical in the axially-symmetric frame. 
For the model with a spherical nucleus we use the radius of 5~km, a nucleus mass of 2.7$\times$10$^{14}$ kg, the total gas production rate (5.1-5.5)$\times$10$^{29}$ s$^{-1}$, and Af$\rho$ of $28,000$~cm \citep[][]{Fink1994,Schleicher1998}. 
The other parameters have been left as free parameters which results in 84 parameter variations. 

\begin{figure}[h]
	\includegraphics[width=\columnwidth]{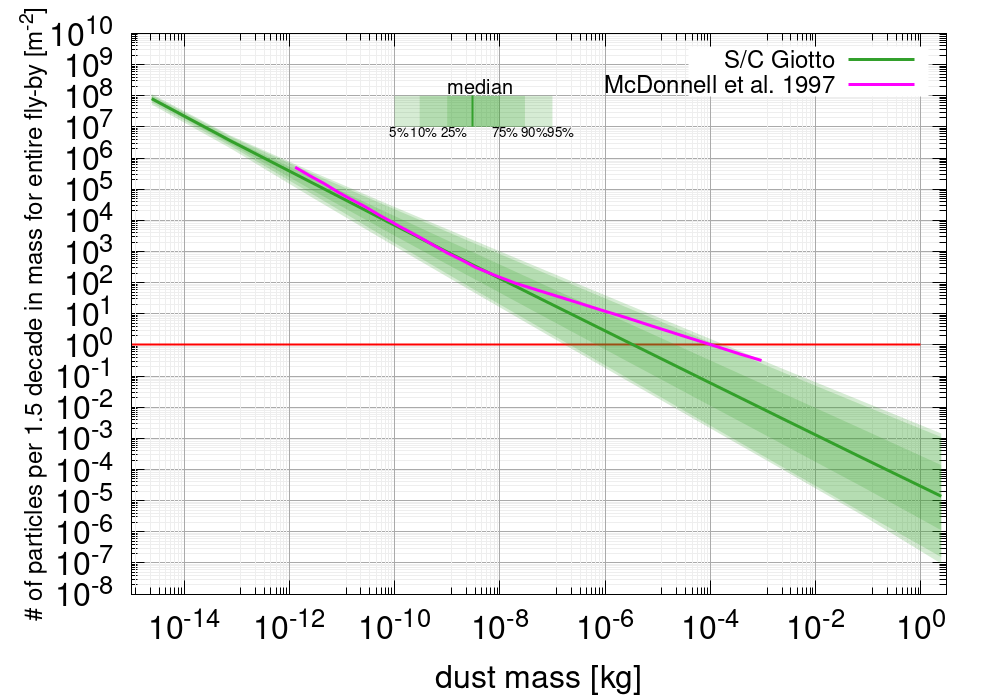}
	\caption{Total number of particles during encounter as a function of mass for the EDCM Halley case and the data by \cite{McDonnell1997}}
	\label{fig:halley-results} 
\end{figure}

The results of the EDCM for 1P/Halley during the Giotto fly-by are shown in Fig.~\ref{fig:halley-results}.
The data for dust masses below $10^{-12}$~kg is an lower limit measurement and should thus not be considered in the comparison. 
Apart from that the data lies within the confidence interval of the model for most of the mass range. 
In the mass range from $10^{-12}$ to $10^{-8}$~kg the model matches well the slope of the Giotto data. 
Because the size distribution for Halley has a break at $\sim3\times10^{-8}$~kg our model cannot match that behaviour as it assumes an unbroken power law. 
But such a power law is implicitly contained in the model because a variety of slopes are simulated and thus are reflected in the uncertainty.

%%%%%%%%%%%%%%%%%%%%%%%%%%%%%%%%%%%%%%%%%%%%%%%%%%%%%%%%%%%%%%%%%%%%%%%%%%
%%%%%%%%%%%%%%%%%%%%%%%%%%%%%%%%%%%%%%%%%%%%%%%%%%%%%%%%%%%%%%%%%%%%%%%%%%
%%%%%%%%%%%%%%%%%%%%%%%%%%%%%%%%%%%%%%%%%%%%%%%%%%%%%%%%%%%%%%%%%%%%%%%%%%
%%%%%%%%%%%%%%%%%%%%%%%%%%%%%%%%%%%%%%%%%%%%%%%%%%%%%%%%%%%%%%%%%%%%%%%%%%
%%%%%%%%%%%%%%%%%%%%%%%%%%%%%%%%%%%%%%%%%%%%%%%%%%%%%%%%%%%%%%%%%%%%%%%%%%
%%%%%%%%%%%%%%%%%%%%%%%%%%%%%%%%%%%%%%%%%%%%%%%%%%%%%%%%%%%%%%%%%%%%%%%%%%
\section{Results}
\subsection{Predictions for the Comet Interceptor spacecraft}\label{sec:resultsCometInterceptor}
The different values for the input parameters listed in Table~\ref{tab:parameters} result in $\sim23,000$ self-consistent combinations (i.e. possible cases of the dust coma). 
To arrive at this set of cases each possible combination of parameter space was checked.
E.g. for each nucleus radius, two different nucleus bulk densities; for each of those combinations three different gas emission distributions; for each of those combinations three different night side activity levels; for each of those combinations twelve different dust size distributions; etc.. 
Multiplying all degrees of freedom in Table~\ref{tab:parameters} results in a total of over 9~million combinations. 
But most of these combinations are not self-consistent and were thus discarded, if they violated a specific constraint, such as $\chi>10$ or, e.g., a given dust production rate resulted in larger Af$\rho$ than we consider for this work.

As described in Sec.~\ref{sec:model} we have extracted for each dust size bin the number of particles encountered along the spacecraft trajectory.

\begin{figure*}[h]
	\includegraphics[width=\textwidth]{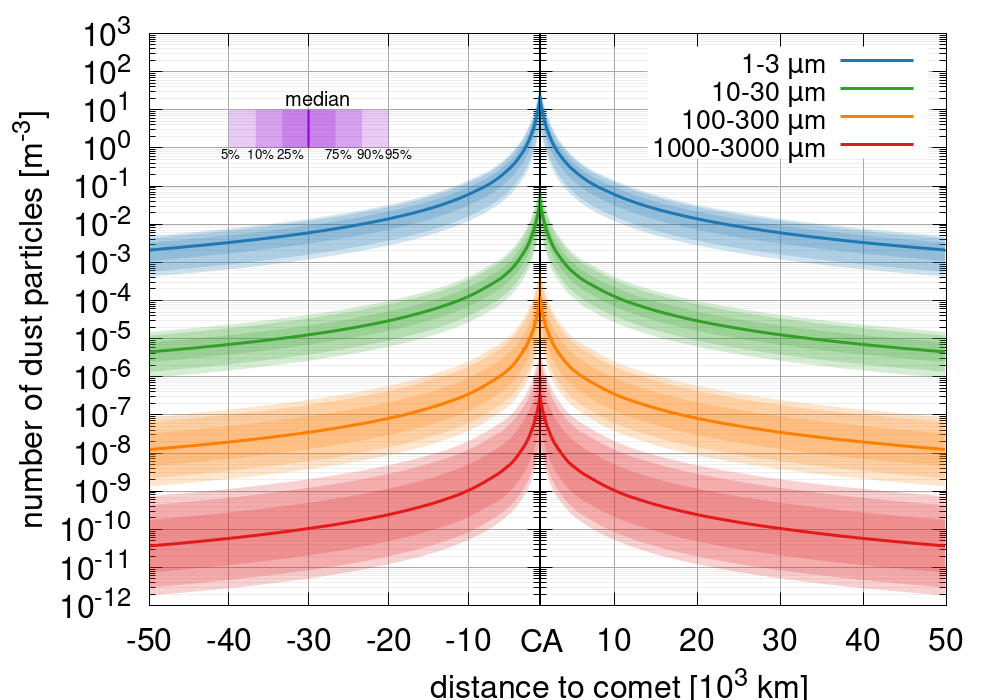}
	\caption{Number of dust particles according to the EDCM along the spacecraft trajectory of spacecraft A as function of cometo-centric distance. The shaded areas show different percentile ranges within which cases fall. Closest approach for S/C A is assumed at 1000~km.}
	\label{fig:numden} 
\end{figure*}

At each point along the trajectory we calculated the median number of particles predicted by the set of all parameter combinations as well as the $5^{th}$, $10^{th}$, $25^{th}$, $75^{th}$, $90^{th}$, and $95^{th}$ percentile.
Each combination within parameter space of the model is assumed to be equally likely.
The results for four size bins and spacecraft A is shown in Fig.~\ref{fig:numden}. 
The shaded areas illustrate the variation in predicted number of particles based on the variation of the input parameters.
These ranges thus reflect to a large degree the uncertainty of our knowledge of the future target of CI.
As the dust size increases the expected number of particles decreases but the uncertainty increases.
Further, the spike in particles around CA highlights that most particles are encountered very close to CA. 
E.g. from cometo-centric distances of 10,000 km to CA at 1,000 km the dust densities increase by roughly 2.5 orders of magnitude.

An important thing to note for the number density is that there does not exist a global median case along the trajectory.
This means what is the median case at far distances may not be the median case at CA.
This also means that the median column density is not equal to the integration of the median number density.
Therefore, we have calculated the percentiles at each distance from the nucleus independently. 

\begin{figure*}[h]
	\includegraphics[width=\textwidth]{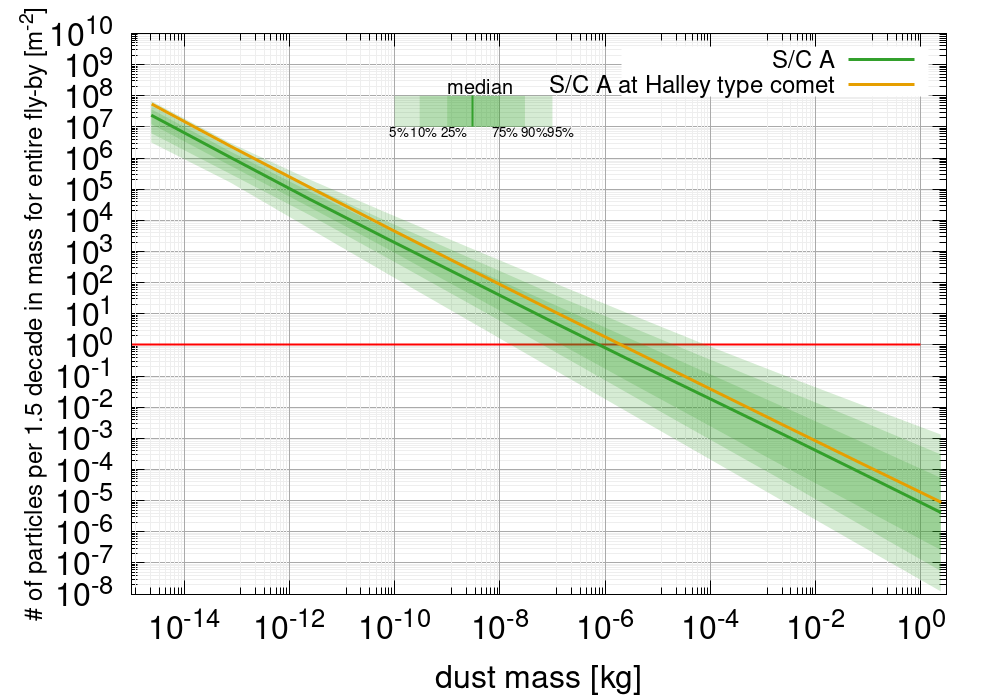}
	\caption{Total number of dust particles encountered according to the EDCM along the spacecraft trajectory of spacecraft A as function dust mass. The shaded areas show different percentile ranges within which cases fall. Additionally the orange curve shows the predicted median densities for a Halley type comet (according to the definition in Sec.~\ref{sec:calibration}).}
	\label{fig:colden} 
\end{figure*}

We have further integrated the total number of particles per square meter expected along the entire fly-by. The results for spacecraft A are shown in Fig.~\ref{fig:colden}. 
The figure shows the number of particles per 1.5 decade in dust mass. 
This means that the value e.g. given at $1~\mu$m or $3$~mm includes the number of particles in the interval from $1-3~\mu$m or $3-10$~mm respectively. 
As another example we expect in the median case $\sim 1$ particle per square meter along the entire fly-by trajectory of spacecraft A in the size range of $300-1000~\mu$m.

Additionally, Fig.~\ref{fig:colden} indicates in orange the median case for an encounter of spacecraft A at comet Halley (as defined Sec.~\ref{sec:calibration}).
The main reason that the median case for spacecraft A lies below a Halley encounter is that the EDCM model assumes a larger range of possible coma environments which turns out to slightly skew to scenarios that produce dust comae with lower dust densities than Halley exhibited.
In a simplified picture, the median target comet in the EDCM is slightly less active than Halley.

Finally, it is worth mentioning that in contrast to the number density there do exist global percentile cases (e.g. a global median case).
All results for all three CI spacecraft and additional supplementary results are available online\footnote{\url{https://www.doi.org/10.5281/zenodo.6906815}}.

%%%%%%%%%%%%%%%%%%%%%%%%%%%%%%%%%%%%%%%%%%%%%%%%%%%%%%%%%%%%%%%%%%%%%%%%%%
%%%%%%%%%%%%%%%%%%%%%%%%%%%%%%%%%%%%%%%%%%%%%%%%%%%%%%%%%%%%%%%%%%%%%%%%%%
%%%%%%%%%%%%%%%%%%%%%%%%%%%%%%%%%%%%%%%%%%%%%%%%%%%%%%%%%%%%%%%%%%%%%%%%%%
%%%%%%%%%%%%%%%%%%%%%%%%%%%%%%%%%%%%%%%%%%%%%%%%%%%%%%%%%%%%%%%%%%%%%%%%%%
%%%%%%%%%%%%%%%%%%%%%%%%%%%%%%%%%%%%%%%%%%%%%%%%%%%%%%%%%%%%%%%%%%%%%%%%%%
%%%%%%%%%%%%%%%%%%%%%%%%%%%%%%%%%%%%%%%%%%%%%%%%%%%%%%%%%%%%%%%%%%%%%%%%%%
\subsection{Predictive strength of parameters}\label{sec:predictiveStrenght}
Based on the ensemble of self-consistent cometary comae described in Sec.~\ref{sec:resultsCometInterceptor} we can explore which parameters are predictive of the dust densities within the coma.
Generally speaking, the parameters we have found to be most predictive of the fly-by column density are the combination of: Af$\rho$, the power law exponent of the dust size distribution, $\beta$, and the dust production rate.
Importantly the predictive power of these three parameters is not equal for all dust sizes.

\begin{figure*}[h]
	\includegraphics[width=\textwidth]{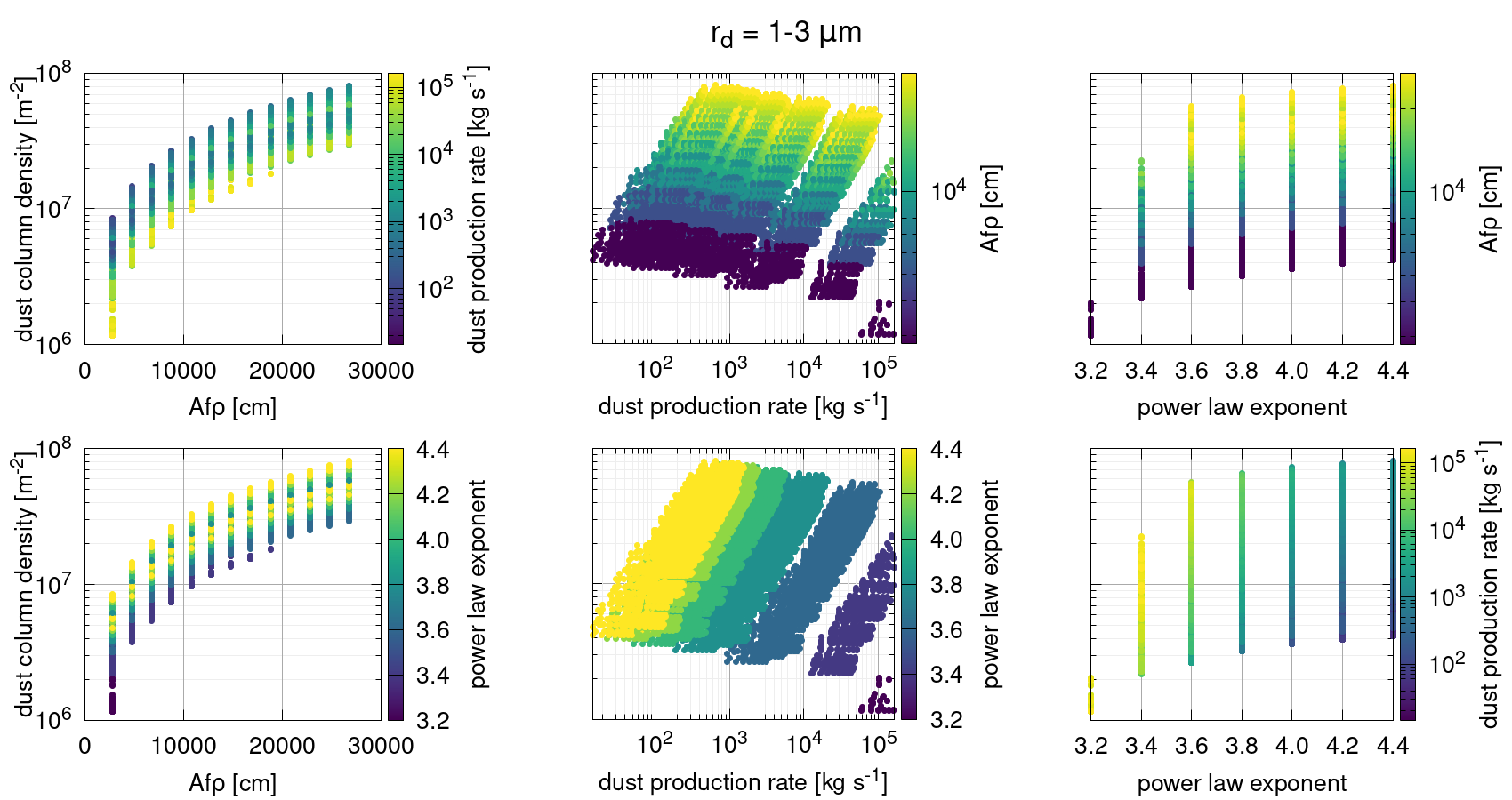}
	\caption{The dust column density for dust particles with radii between $1-3~\mu$m at the Comet Interceptor spacecraft A is shown as a function of different combinations of Af$\rho$, the power law exponent, $\beta$, and the dust production rate.}
	\label{fig:correlationSmallParticles} 
\end{figure*}
Figure~\ref{fig:correlationSmallParticles} shows the dust column density for dust particles with radii between $1-3~\mu$m at the Comet Interceptor spacecraft A for different combinations of Af$\rho$, the power law exponent, $\beta$, and the dust production rate.
Af$\rho$ by itself has the highest predictive power in this case.
The dust column density for a given Af$\rho$ is constrained to $\sim$half an order of magnitude in all cases except for the lowest Af$\rho$ (left panels in Fig.~\ref{fig:correlationSmallParticles}).
For any given dust production rate (center panels in Fig.~\ref{fig:correlationSmallParticles}) or power law exponent (right panels in Fig.~\ref{fig:correlationSmallParticles}) by themselves the possible column densities span at least one order of magnitude in most cases.
Af$\rho$ in combination with either the power law index or the dust production rate will tightly constrain the column density.

\begin{figure*}[h]
	\includegraphics[width=\textwidth]{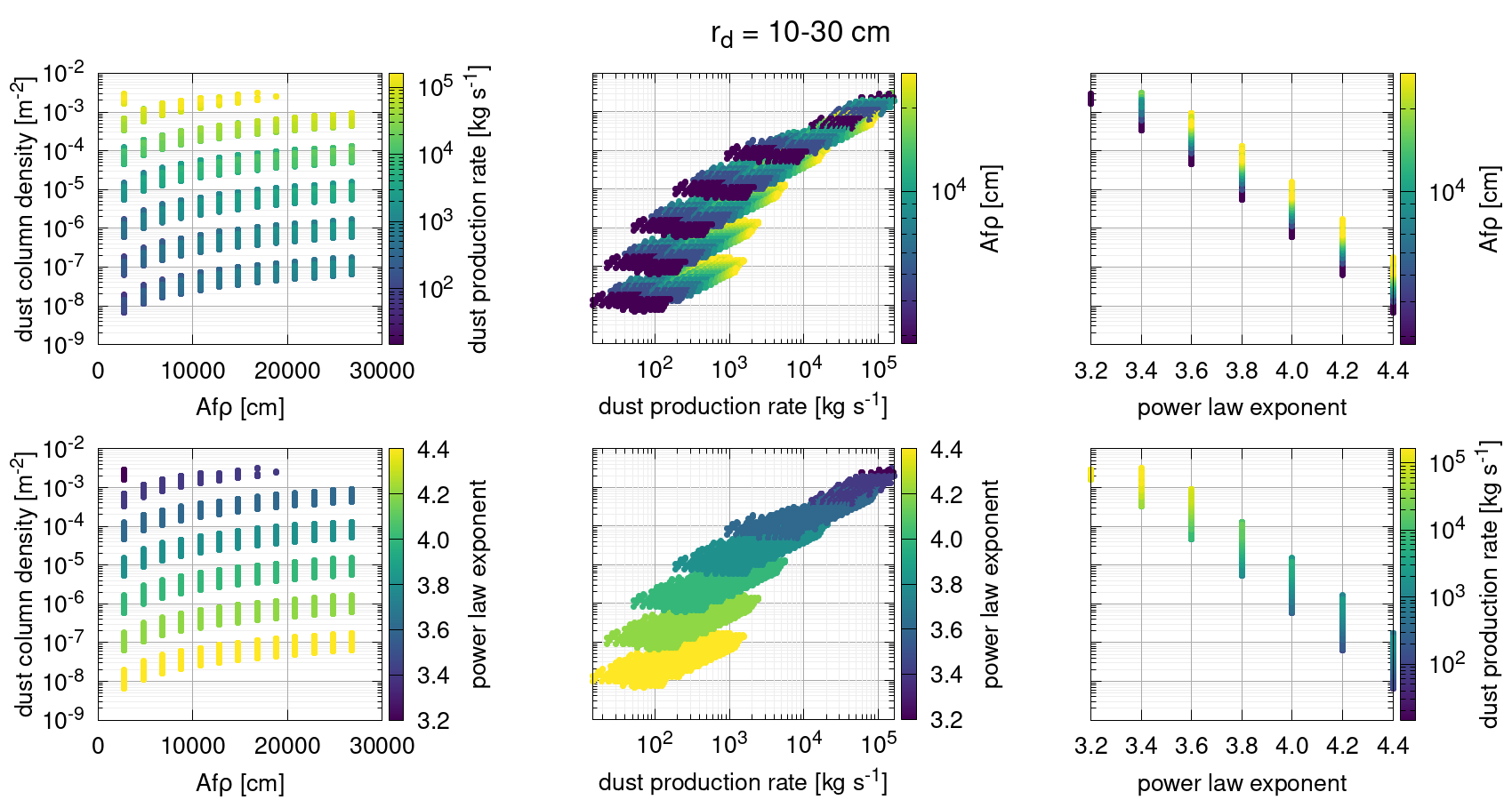}
	\caption{The dust column density for dust particles with radii between $10-30~$cm at the Comet Interceptor spacecraft A is shown as a function of different combinations of Af$\rho$, the power law exponent, $\beta$, and the dust production rate.}
	\label{fig:correlationLargeParticles} 
\end{figure*}

The behaviour for large particles (e.g. 10-30 cm) is very different from the one described above for small particles.
In this case Af$\rho$ is the most unpredictive of the three parameters (left panels Fig.~\ref{fig:correlationLargeParticles}).
For a given Af$\rho$ the possible dust column densities span $\sim$5 orders of magnitude.
For large particles the power law index is generally the most predictive parameter.
For a given $\beta$ alone (right panels in Fig.~\ref{fig:correlationLargeParticles}) the expected dust column densities can be predicted to $\sim$1.5 orders of magnitude accuracy.
The dust production rate by itself (center panels in Fig.~\ref{fig:correlationLargeParticles}) has variable predictive power.
The predicted range of dust column densities is largest at small dust production rates and then decreases with increasing dust production rates.
The best predictions of the dust column density of large particles can be achieved with a combination of the power law exponent and either Af$\rho$, or the dust production rate (left panels Fig.~\ref{fig:correlationLargeParticles}).
By adding one of these constraints to the power law exponent the predicted dust column densities can be determined down to an accuracy of $\sim$half an order of magnitude.

\begin{figure}[h]
	\includegraphics[width=\columnwidth]{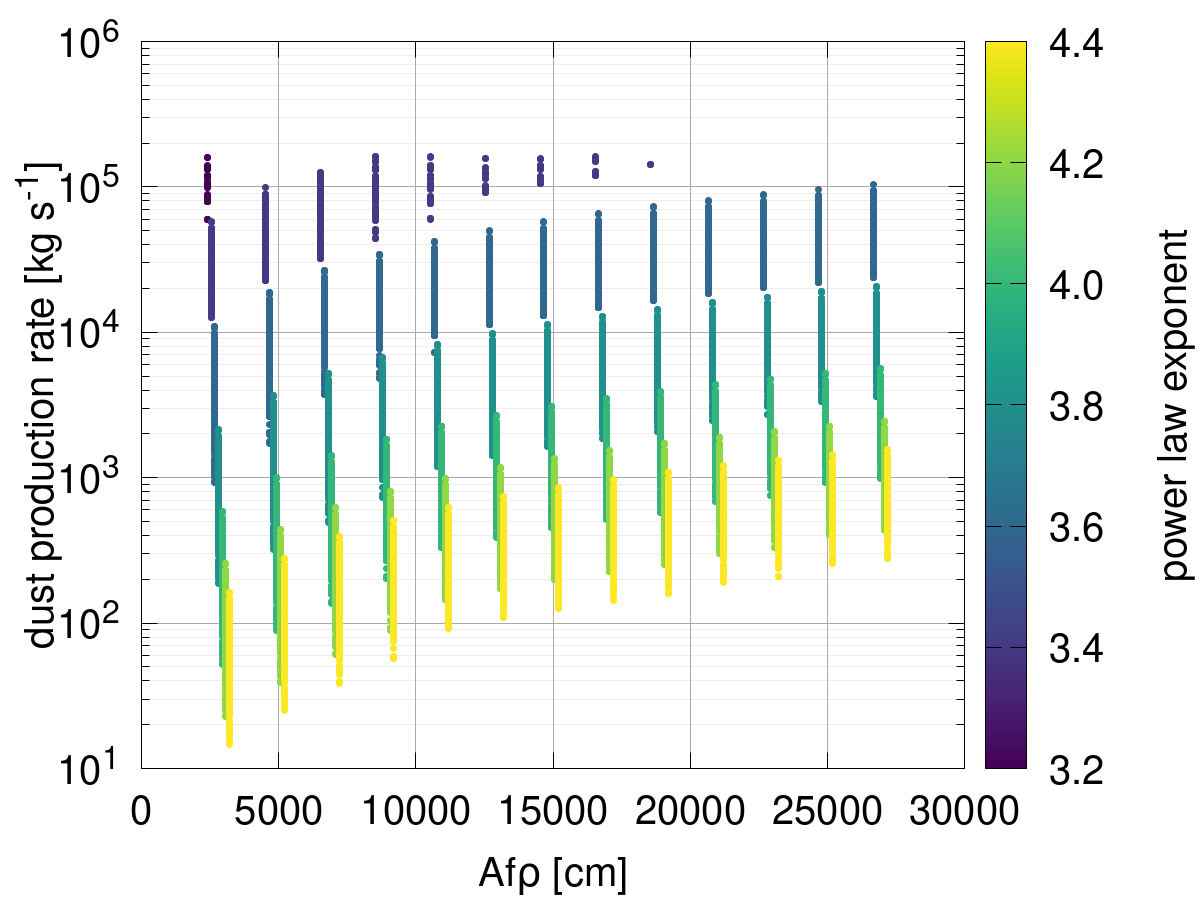}
	\caption{The dust production rate as a function of Af$\rho$ and the power law exponent, $\beta$, of the dust size distribution. The values of Af$\rho$ are slightly offset depending on the power law exponent to show the overlap resulting from $\beta$. Af$\rho$ by itself is a poor predicter of the dust production rate of a comet.}
	\label{fig:Afrho-to-Qd} 
\end{figure}
Often Af$\rho$ is used to estimate the dust production rate of a comet (e.g. Af$\rho$[cm] $\sim$ Q[kg/s], \citet{AHearn1995}).
With our ensemble of dust comae we can test how large the associated uncertainties are.
Figure~\ref{fig:Afrho-to-Qd} shows the production rate as a function of Af$\rho$ and the power law index.
First, it is quite clear that Af$\rho$ is a rather poor preditor of the dust production rate.
For a given Af$\rho$ the dust production rate spans $\sim$2.5 to $\sim$4 orders of magnitude.
This illustrates why Af$\rho$ by itself should not be used to estimate a comet's dust production rate.
The estimates of the dust production rate can however be greatly improved when the the power law index of the size distribution is known.
In this case the uncertainty in the dust production rate can be reduced to $\sim$1 order of magnitude.

\begin{figure*}[h]
	\includegraphics[width=\textwidth]{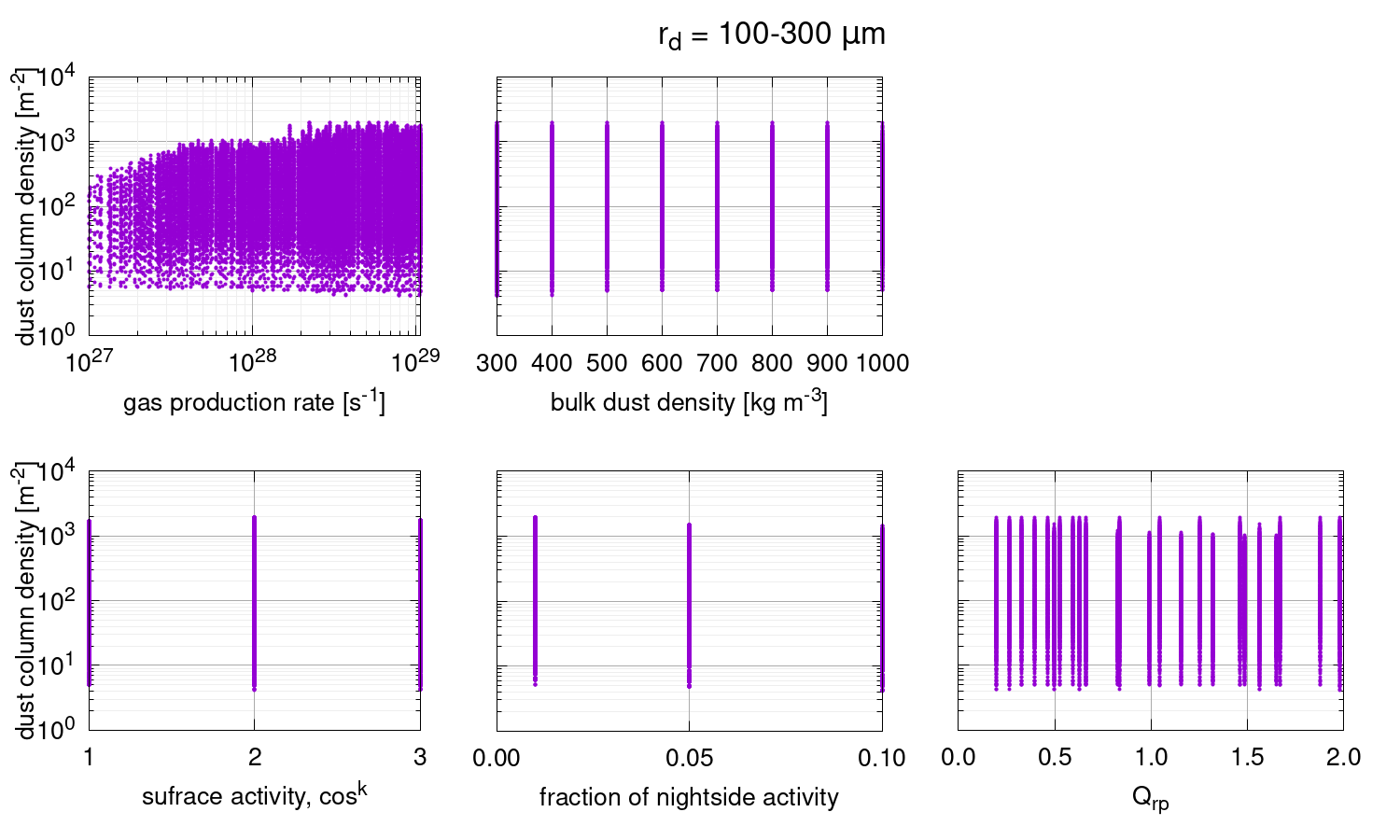}
	\caption{The dust column density for dust particles with radii between $100-300~\mu$m at the Comet Interceptor spacecraft A is shown as a function of different combinations of the gas production rate, dust bulk density, surface activity distribution, night time activity, and radiation pressure efficiency. These quantities are not predictive of the dust column density.}
	\label{fig:uncorrelated} 
\end{figure*}
Finally, for completeness we show in Fig.~\ref{fig:uncorrelated} that the other parameters in the model are not predictive of the dust column density.
There are some minor trends e.g. in the fraction of night side activity (bottom centre panel of Fig.~\ref{fig:uncorrelated}).
In this case this minor correlation is easily understood.
The column density considered here is located on the day side (i.e. where the CI fly-by of spacecraft A will occur).
Therefore, naturally as more activity occurs on the night side the relative amount of dust on the day side will decrease.

%%%%%%%%%%%%%%%%%%%%%%%%%%%%%%%%%%%%%%%%%%%%%%%%%%%%%%%%%%%%%%%%%%%%%%%%%%
%%%%%%%%%%%%%%%%%%%%%%%%%%%%%%%%%%%%%%%%%%%%%%%%%%%%%%%%%%%%%%%%%%%%%%%%%%
%%%%%%%%%%%%%%%%%%%%%%%%%%%%%%%%%%%%%%%%%%%%%%%%%%%%%%%%%%%%%%%%%%%%%%%%%%
%%%%%%%%%%%%%%%%%%%%%%%%%%%%%%%%%%%%%%%%%%%%%%%%%%%%%%%%%%%%%%%%%%%%%%%%%%
%%%%%%%%%%%%%%%%%%%%%%%%%%%%%%%%%%%%%%%%%%%%%%%%%%%%%%%%%%%%%%%%%%%%%%%%%%
%%%%%%%%%%%%%%%%%%%%%%%%%%%%%%%%%%%%%%%%%%%%%%%%%%%%%%%%%%%%%%%%%%%%%%%%%%
\subsection{Dust speeds and impact angles}\label{sec:dustSpeeds}
\begin{figure*}[h]
	\includegraphics[width=\textwidth]{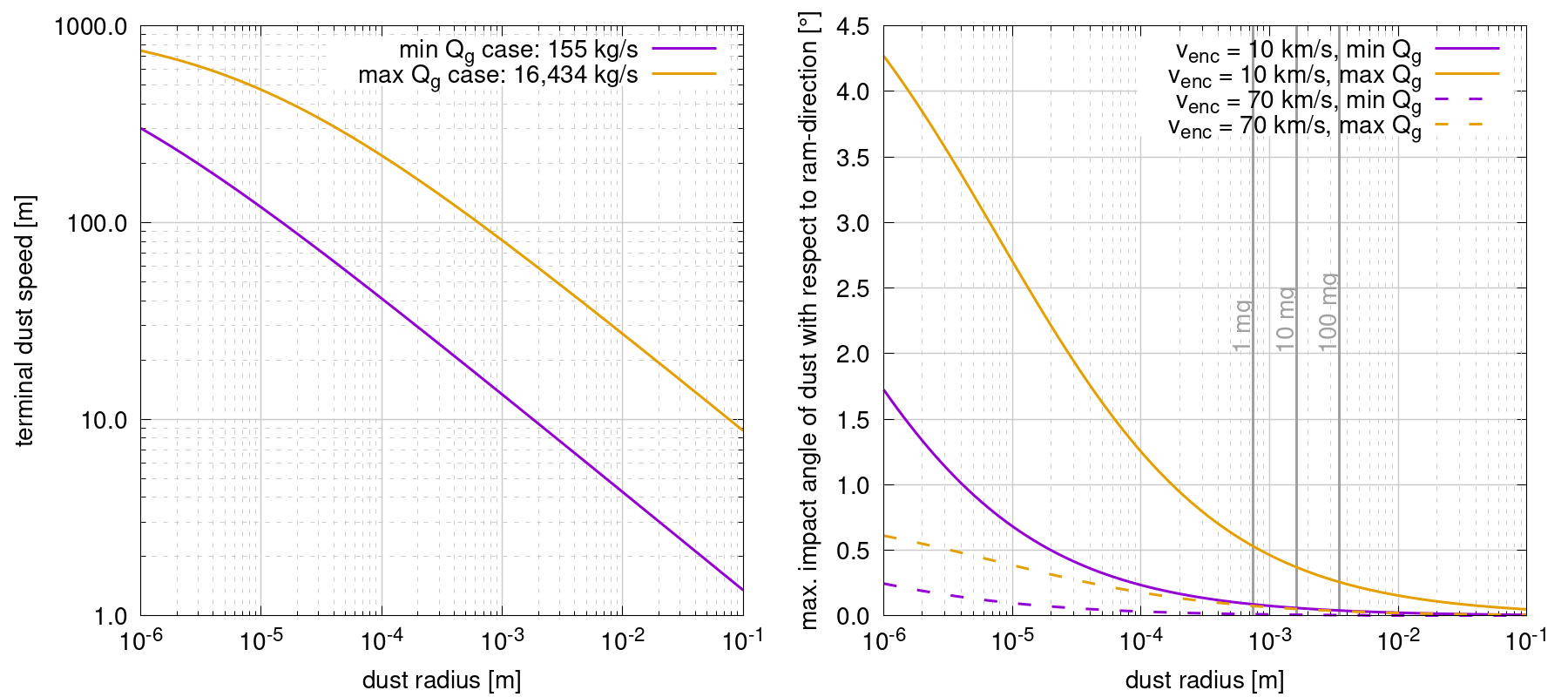}
	\caption{The left panel shows the terminal dust speed, $v_{d,term}$, according to Eq.~\ref{eq:termDustSpeed} for two gas cases. The lowest gas production rate case considered in the EDCM in purple (min $Q_g$ case, where $R_N=2$~km and $Q_g=155$~kg/s) and the highest gas production rate case in orange (max $Q_g$ case, where $R_N=5$~km and $Q_g=16,434$~kg/s). The right panel shows the maximum angle of impact of dust particles with respect to the ram-direction of the spacecraft. The solid lines are for a fly-by speed of $10$~km/s and the dashed lines for $70$~km/s. The colours are the min. and max. gas production cases as in the left panel. The corresponding radii of particles with masses 1~mg, 10~mg, and 100~mg are indicated by the grey lines. }
	\label{fig:termDustSpeed} 
\end{figure*}

We would be remiss not to comment on the dust speeds and what those would mean for the impact angle on the spacecraft.
With a simple estimates we want to illustrate that the spacecraft motion with respect to the comet will dominate the flux direction of the dust onto the spacecraft in almost all cases.
The terminal dust speeds cannot exceed the gas speed with an upper limit of $\sim 1,000$~m/s.
In the case of isotropic gas expansion from a spherical nucleus the terminal dust velocity, $v_{d,term}$, can be calculated from the approximation of the numerical solution:
\begin{equation}\label{eq:termDustSpeed}
v_{d,term} = \frac{58.903 \sqrt{T_N}}{1+0.6 \sqrt{1.5/{\rm Iv}}} \quad,
\end{equation}
where
\begin{equation}
{\rm Iv} = 4.038 \cdot 10^{-29} \frac{Q_g}{R_N \varrho_d a \sqrt{T_N}} \quad,
\end{equation}
and where $T_N$ is the nucleus temperature (to be set to 300~K for our purposes here), $Q_g$ is the gas production rate [molecules/s], $R_N$ is the radius of the nucleus [m], and $\varrho_d$ is the bulk dust density
.
The left panel of Fig.~\ref{fig:termDustSpeed} shows the terminal dust speeds for the maximum and minimum gas production rates considered in the EDCM for CI.
Particles larger than $\sim600~\mu$m are always slower than $100$~m/s.

From the terminal dust speed we can derive the maximum impact angle of a dust particle with respect to the ram-direction of the spacecraft given a certain fly-by velocity.
Currently fly-by velocities between 10~km/s and 70~km/s are considered for CI.
The most likely fly-by speed is expected to be 50~km/s.

The right panel of Fig.~\ref{fig:termDustSpeed} shows this angle which is always smaller than $\sim0.5^{\circ}$ for particles with masses larger than 1~mg.
The maximum impact angle occurs for the smallest particles ($1~\mu$m) in the highest gas production rate case and is $\sim4.3^{\circ}$.
This illustrates that the impact angles is dominated by the spacecraft fly-by velocity.

That such impacts can nevertheless cause some issues for attitude control is shown in \citet{Haslebacher2022}.

%%%%%%%%%%%%%%%%%%%%%%%%%%%%%%%%%%%%%%%%%%%%%%%%%%%%%%%%%%%%%%%%%%%%%%%%%%
%%%%%%%%%%%%%%%%%%%%%%%%%%%%%%%%%%%%%%%%%%%%%%%%%%%%%%%%%%%%%%%%%%%%%%%%%%
%%%%%%%%%%%%%%%%%%%%%%%%%%%%%%%%%%%%%%%%%%%%%%%%%%%%%%%%%%%%%%%%%%%%%%%%%%
%%%%%%%%%%%%%%%%%%%%%%%%%%%%%%%%%%%%%%%%%%%%%%%%%%%%%%%%%%%%%%%%%%%%%%%%%%
%%%%%%%%%%%%%%%%%%%%%%%%%%%%%%%%%%%%%%%%%%%%%%%%%%%%%%%%%%%%%%%%%%%%%%%%%%
%%%%%%%%%%%%%%%%%%%%%%%%%%%%%%%%%%%%%%%%%%%%%%%%%%%%%%%%%%%%%%%%%%%%%%%%%%
\section{Guiding mission design and safety}\label{sec:impact}
In the introduction, we highlighted five areas that motivate this modelling approach.
Here we revisit these points to show how the EDCM is/will be used to guide mission design and safety.

\subsection{Trajectory design and spacecraft shielding}
The choices of spacecraft trajectory and spacecraft shielding are interwoven.
A more distant closest approach distance or an encounter through the anti-solar point rather than the sub-solar point would result in fewer particle impacts.
How much the dust densities can be reduced through changes in the fly-by trajectory is detailed in Sec.~\ref{sec:scalingResults}.
Therefore, there is a trade-off between the amount of shielding needed and how close the spacecraft can fly to the comet.
The CI spacecraft are designed to survive a Halley/Giotto type comet encounter.
This corresponds to the median case of the EDCM. 
The model is being used to assess the risk of particle hits, in particular the number of particles in the different size bins.
The shielding of the CI spacecraft have been sized following this criterion. 

\subsection{
Performance of scientific instruments and star trackers}
One of the issues during operations of Rosetta at comet 67P was triggered by star
trackers falsely identifying dust particles as stars \citep{Accomazzo2017AcAau}.
Modern star trackers in combination with the EDCM dust densities will allow for in-depth bench-marking and thus safe operations. 

Further, the dust accumulation rate, in the ram-direction, can be calculated using the column densities of the model.
This is particularly important for instruments such as the Dust Impact Sensor and Counter (DISC) which are placed on spacecraft A and probe B2.
DISC will measure the momentum and mass of the particles when they impact its sensitive surface.

\subsection{Attitude control}
The impact of dust particles onto the spacecraft's surface will impact attitude control and thus the pointing of the instruments.
As shown by \citet{Haslebacher2022} this can be significant and is a potential concern e.g. for keeping the nucleus within the field of view of the COmet CAmera (CoCa) on spacecraft A.
They found that a total change in angular velocity that needs to be corrected can reach $1-10^{\circ}$/s resulting in a median shift of the target object on the CoCa detector of more than 10 pixels.
\citet{Haslebacher2022} used an idealised coma model and thus the updated values of the EDCM will allow for a refined assessment of this issue and thus provide better information on the need and scope of mitigating strategies.

%%%%%%%%%%%%%%%%%%%%%%%%%%%%%%%%%%%%%%%%%%%%%%%%%%%%%%%%%%%%%%%%%%%%%%%%%%
%%%%%%%%%%%%%%%%%%%%%%%%%%%%%%%%%%%%%%%%%%%%%%%%%%%%%%%%%%%%%%%%%%%%%%%%%%
%%%%%%%%%%%%%%%%%%%%%%%%%%%%%%%%%%%%%%%%%%%%%%%%%%%%%%%%%%%%%%%%%%%%%%%%%%
%%%%%%%%%%%%%%%%%%%%%%%%%%%%%%%%%%%%%%%%%%%%%%%%%%%%%%%%%%%%%%%%%%%%%%%%%%
%%%%%%%%%%%%%%%%%%%%%%%%%%%%%%%%%%%%%%%%%%%%%%%%%%%%%%%%%%%%%%%%%%%%%%%%%%
%%%%%%%%%%%%%%%%%%%%%%%%%%%%%%%%%%%%%%%%%%%%%%%%%%%%%%%%%%%%%%%%%%%%%%%%%%
\section{Scaling the results}\label{sec:scalingResults}
The nominal trajectories (table~\ref{tab:parameters}) of the model only vary in CA distance but not in the two angles $\alpha_T$ and $\beta_T$ ($\alpha$ and $\beta$ in Fig.~\ref{fig:giotto_trj}) which are both assumed to be zero.
Because this will not be the case for all three spacecrafts of CI we have tested the model to the sensitivity of changes in $\alpha_T$, $\beta_T$, as well as the CA distance, $r_{CA}$.
Furthermore, the results might be needed for a different dust size or size interval than we have used in this work.

\subsection{Variations in CA distance}
First, we explored variations in the CA distance. 
For a force free, radially expanding coma the column density scales with $1/$CA.
\begin{figure}[h]
	\includegraphics[width=\columnwidth]{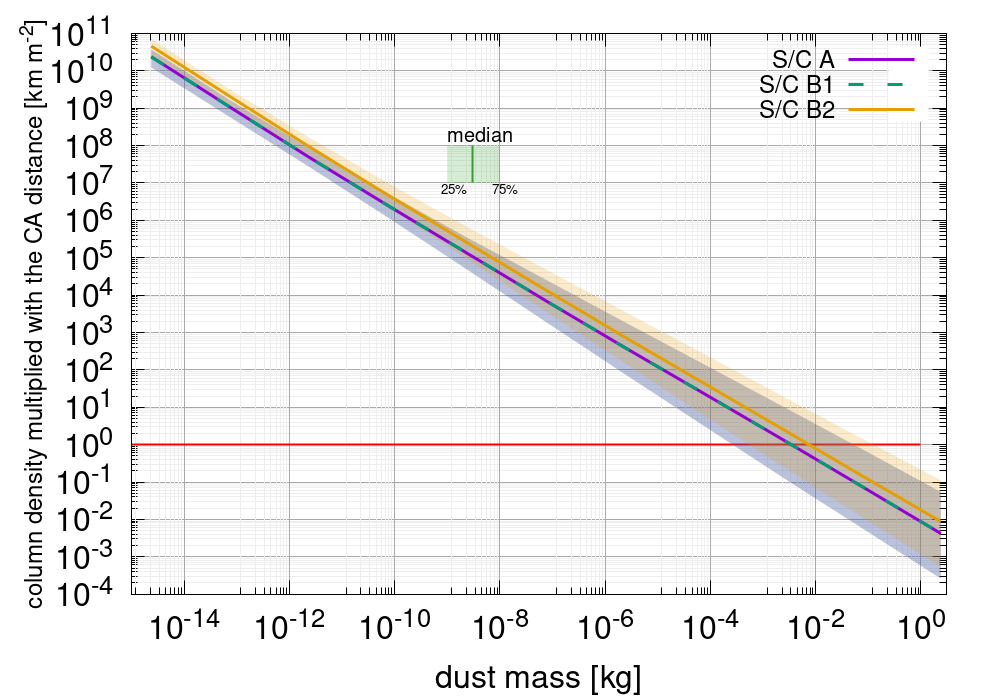}
	\caption{Column density multiplied with the CA distance as a function of dust mass.}
	\label{fig:sensitivity-CA} 
\end{figure}
Or put another way, the product of the column density and the CA distance is constant. 
Fig.~\ref{fig:sensitivity-CA} shows this for the three S/C and reveals that this scaling works well for S/C A \& B1 but not perfectly for S/C B2 which results in densities roughly a factor of 2 higher than expected from that scaling of the other two S/C. 
This is likely due to the fact that S/C B2 gets very close to the nucleus (down to $20 R_N$ in some of the cases considered). 

In a sense S/C B1 is the nominal case as it is not too close to the nucleus and not too far from the nucleus (where we need to extrapolate the dust results from our modelling domain and thus "force" a $1/$CA behaviour).
The medians of the different S/C are all still within the centre 50 percentiles of the cases. 
Thus the uncertainty from the unknown parameters of the model is much larger than the uncertainty of this simple scaling. 

Also the assumed differences compared here are very large, therefore for small changes in the CA distance the results of the model can be scaled with
\begin{equation}
    N_{col}(r_{CA}) = N_{col}(r_{CA; 0}) \frac{r_{CA; 0}}{r_{CA}}\quad,
\end{equation}
where $r_{CA; 0}$ is the reference CA distance.
The number densities, $N_{num}$, scale respectively with the square of the closest approach distance, i.e.
\begin{equation}
    N_{num}(r_{CA}) = N_{num}(r_{CA; 0}) \left( \frac{r_{CA; 0}}{r_{CA}} \right)^2   \quad.
\end{equation}

\subsection{Variations in $\alpha_T$ and $\beta_T$}
Next we explore changes in $\alpha_T$ ($\alpha$ in Fig.~\ref{fig:giotto_trj}).
Changes to only $\alpha_T$ ($\beta_T=0^{\circ}$) merely change the phase angle at CA but retain a symmetric trajectory around CA. 
We've run the EDCM for $\alpha_T=0^{\circ}, 15^{\circ}, 30^{\circ}, 45^{\circ}, 60^{\circ}$ for S/C B2. 
\begin{figure*}
\hfill
\subfigure[normalised column density as a function of $\alpha_T$ ($^{\circ}$)]{\includegraphics[width=0.49\linewidth]{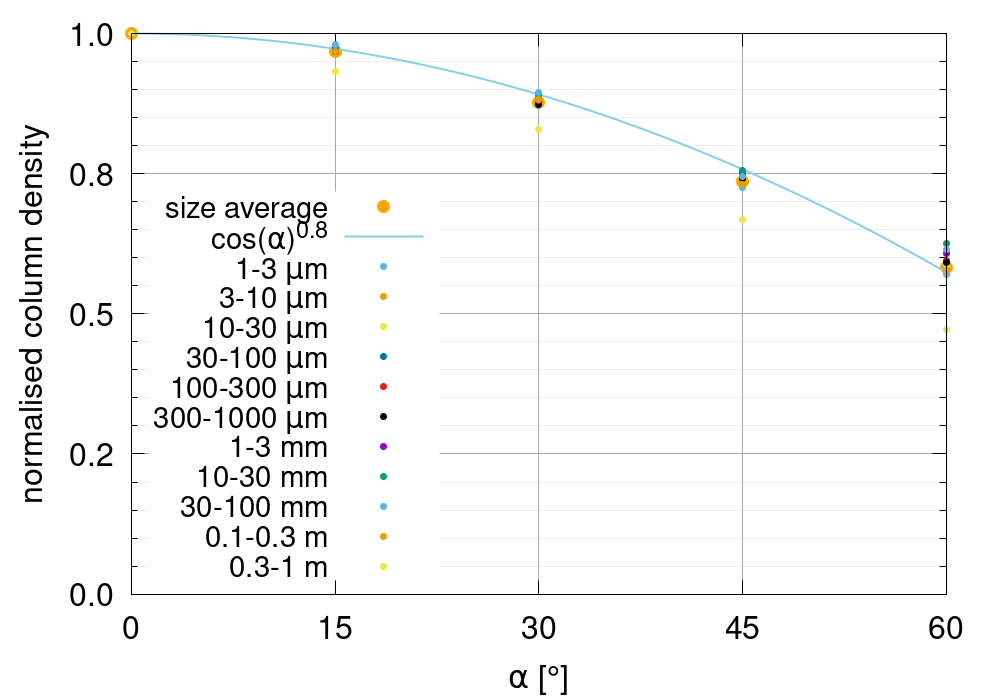}}
\hfill
\subfigure[normalised column density as a function of $\beta_T$ ($^{\circ}$)]{\includegraphics[width=0.49\linewidth]{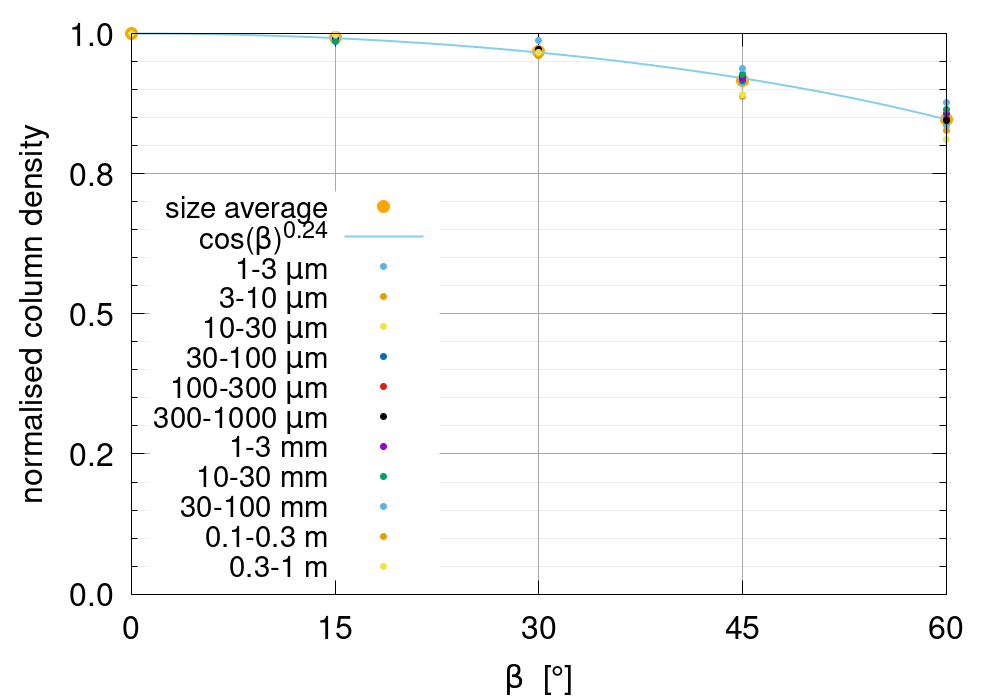}}
\hfill
\caption{Dependence of the column density with $\alpha_T$ and $\beta_T$. $\alpha_T$ is the angle between the orbital plane of the comet and the direction from the comet to the spacecraft at closet approach ($\alpha$ in Fig.~\ref{fig:giotto_trj}). $\beta_T$ is the angle between the direction of the Sun and the direction from the nucleus to the closest approach projected into the orbital plane ($\beta$ in Fig.~\ref{fig:giotto_trj} and not to be confused with $\beta$ the power law). }
\label{fig:sensitivity-alpha}
\end{figure*}
Figure~\ref{fig:sensitivity-alpha}~a) shows the normalised column density to the values of each dust sizes column density at $\alpha_T=0^{\circ}$ as well as the size averaged relative column density. 
The column densities scale roughly as a function of the cosine but barely reach a factor of 2 decrease at $\alpha_T=60^{\circ}$ which again is much less than the uncertainties inherent to the EDCM from the large uncertainties of the input parameters. 
We can therefore use the rough empirical relation
\begin{equation}
    N_{col}(\alpha_T) = N_{col}(0) \cos^{0.8}(\alpha_T)
\end{equation}
to scale the nominal results to different values of $\alpha_T$.

\begin{figure*}
\hfill
\subfigure[$\alpha_T=\beta_T=0^{\circ}$]{\includegraphics[width=0.49\linewidth]{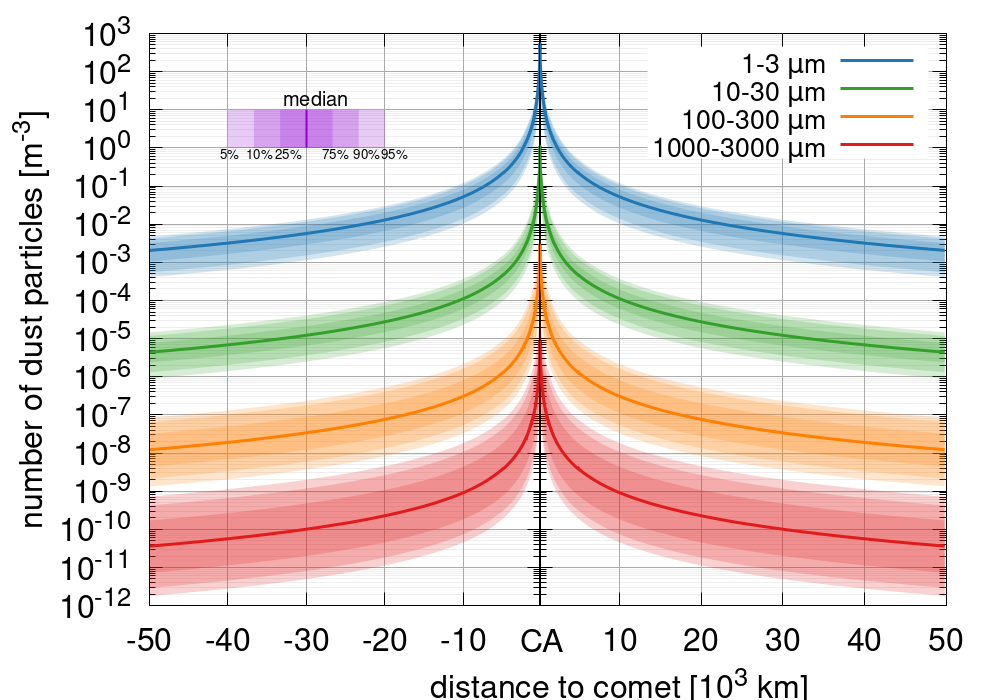}}
\hfill
\subfigure[$\alpha_T=\beta_T=0^{\circ}$; zoom view]{\includegraphics[width=0.49\linewidth]{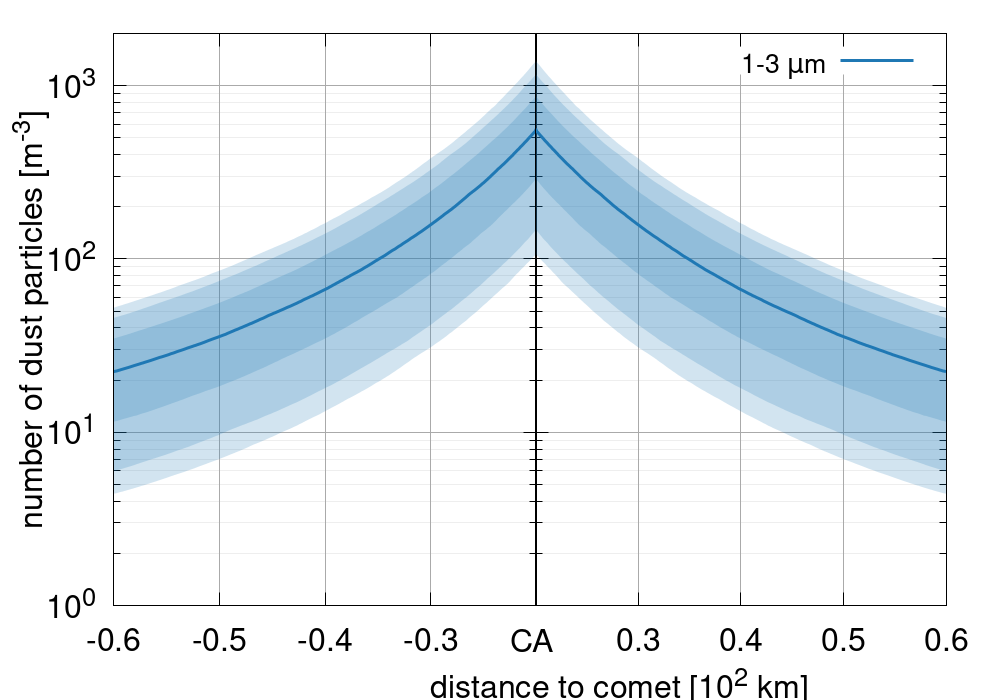}}
\hfill
\subfigure[$\alpha_T=0^{\circ}; \beta_T=60^{\circ}$]{\includegraphics[width=0.49\linewidth]{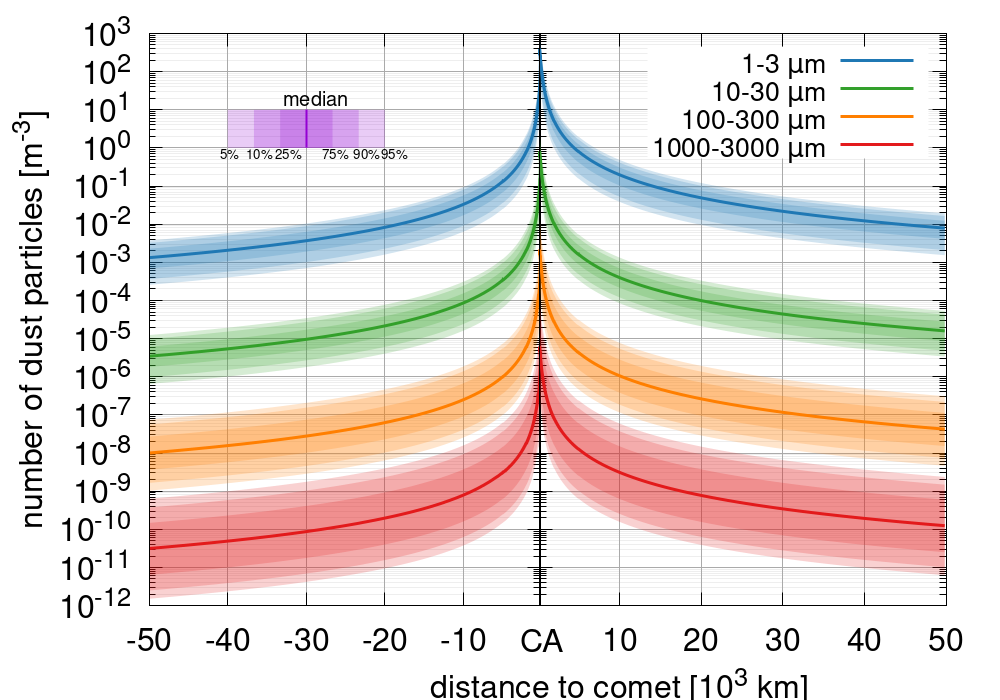}}
\hfill
\subfigure[$\alpha_T=0^{\circ}; \beta_T=60^{\circ}$; zoom view]{\includegraphics[width=0.49\linewidth]{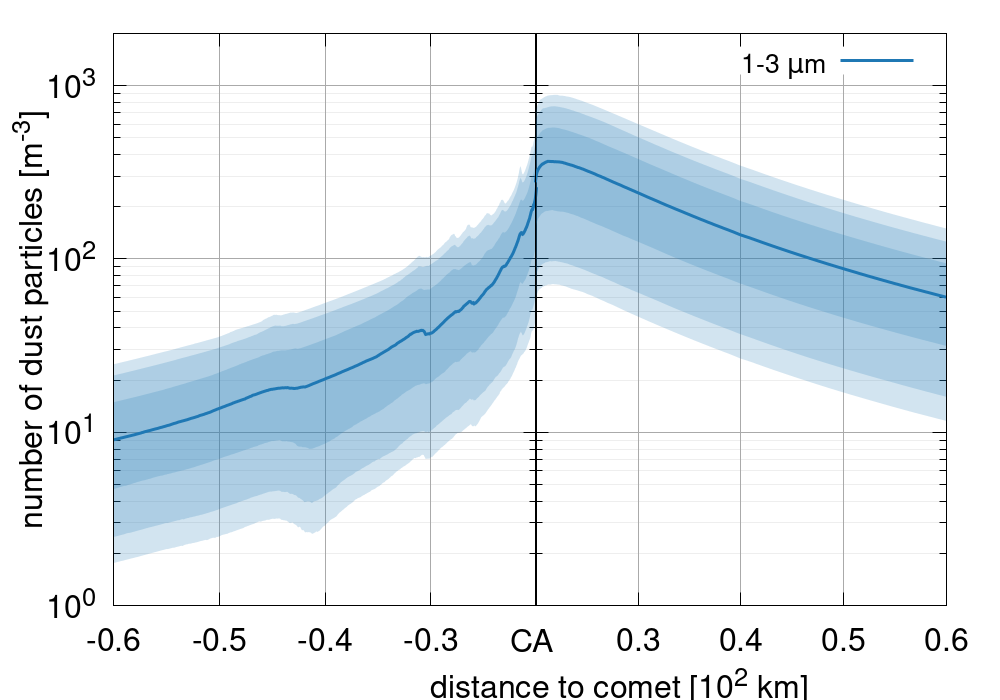}}
\hfill
\caption{Number densities as a function of cometo centric distance for S/C B2  for different dust sizes.}
\label{fig:sensitivity-beta}
\end{figure*}

Finally, we varied $\beta_T$ ($\beta$ in Fig.~\ref{fig:giotto_trj} and not to be confused with $\beta$ the power law) which not only changes the phase angle at CA to $\beta_T$ (if $\alpha_T=0^{\circ}$) but also skews the trajectory such that the asymptotic phase angles are $90\pm\beta_T$. 
If $\alpha_T=0^{\circ}$, as assumed here, the S/C will usually go through zero phase shortly after/before CA. 
Figure~\ref{fig:sensitivity-alpha}~b) shows the normalised column density to the values of each dust sizes column density at $\beta_T=0^{\circ}$ as well as the size averaged relative column density. 
The dependency of the column density on $\beta_T$ is weaker than that of $\alpha_T$. The maximum decrease,at $\beta_T=60^{\circ}$, is $\sim15\%$. We can use the rough empirical relation
\begin{equation}
    N_{col}(\beta_T) = N_{col}(0) \cos^{0.24}(\beta_T)
\end{equation}
to scale the nominal results to different values of $\beta_T$.

But the change in $\beta_T$ will result in an asymmetric dust distribution along the S/C trajectory. 
This is illustrated for S/C B2 in Fig.~\ref{fig:sensitivity-beta} showing the examples where $\beta_T$ is $0^{\circ}$ (top row) and $60^{\circ}$ (bottom row).
In the case where $\beta_T=60^{\circ}$ the inbound densities are lower than for the symmetric trajectory because the inbound asymptotic phase angle is $150^{\circ}$ instead of $90^{\circ}$. 
Conversely, the outbound densities are higher than for the symmetric trajectory because the outbound asymptotic phase angle is $30^{\circ}$ instead of $90^{\circ}$. 
Further the maximum number density will occur shortly after CA instead of at CA.

Finally, we should state that all changes in the CA distance, $\alpha_T$, and $\beta_T$ result in small changes in the column densities compared to the inherent uncertainties resulting from our poor knowledge of the input parameters.

\subsection{Scaling of results to other dust sizes}
In the case that the model does not provide the results for the desired dust radius or size of the dust bin (by default half a decade) simple scalings can be applied.
We should point out to use these scaling with caution as they might break down if pushed to their extremes (e.g. to very short size intervals).
All scaling we provide here apply only to the respective median cases of the column density of the respective spacecraft.

First we will define some general functions to calculate the respective scaling later on.
The column density as a function of the dust radius, $a$, follows a double asymptotic power law of the form
\begin{equation}\label{eq:colDenFit}
    N_{col,fit}(a) = C_1 a^{-\beta_1} + C_2 a^{-\beta_2} \quad,
\end{equation}
where $\beta_i$ are the respective asymptotic power law exponents, and 
\begin{equation}\label{eq:powerConst}
    C_i(a_{0,i},N_{0,i},\beta_i) = N_{0,i} a_{0,i}^{\beta_i}
\end{equation}
is the normalisation constant given where $N_{0,i}$ is the column density at reference size $a_{0,i}$.
The values of these parameters for the three spacecraft are given in Table~\ref{tab:fitParameter}.

\begin{table}[h]
\caption{Parameters for interpolation between dust sizes and bin interval lengths.}
\label{tab:fitParameter}
\begin{tabular}{l|cccccc}
 S/C & $a_{0,1}$ & $N_{0,1}$ & $\beta_1$ & $a_{0,2}$ & $N_{0,2}$ & $\beta_2$   \\\hline
 A   &           & 1.5E7     &           &           & 4.1E-6    & \\
 B1  & 1E-6~m    & 3.0E7     & $3$       &  1E-1~m   & 8.2E-6    & 2.48\\
 B2  &           & 1.5E8     &           &           & 4.1E-5    &
\end{tabular}
\end{table}

Next we define the length of our dust size bins as
\begin{equation}\label{eq:binLength}
    \Delta a = 10^{d}\quad,
\end{equation}
where $d$ is a decade fraction, e.g. 0.5 for half a decade or 0.25 for a quarter decade. 
From this it follow that a specific dust bin starting at $a$ will extend to $a\times \Delta a$.

To interpolate the results from one dust bin radius to another while keeping the length of the dust bin ($\Delta a$) unchanged we can proceed in the following manner.
When the target dust size is $a_t$ we lookup the dust bin size, $a_r$, which is just smaller than our target size.
This reference dust bin, $a_r$, has a certain column density, $N_{col}(a_r)$, according to the EDCM.
Using Eq.~\ref{eq:colDenFit} the column density for our target size, $a_t$ (covering the dust range from $a_t$ to $a_t \times \Delta a$), will be
\begin{equation}
    N_{col}(a_t) = N_{col}(a_r) \frac{N_{col,fit}(a_t)}{N_{col,fit}(a_r)} \quad.
\end{equation}

\subsection{Interpolation of dust bin length}
Next we discuss how the length of the dust bins can be changed.
The results of the EDCM are provided for bins with a length of half a decade in dust radius.
If one wants to know the results for a bin size of a quarter decade instead of a half a decade the following procedure can be applied.
We should caution the reader at this point that we have tested this scaling only down to a quarter decade.
While we believe that it also holds for smaller intervals the user should be aware of this limitation and that smaller bin lengths could be associated with additional uncertainty.

The provided solution have dust bin lengths which are half a decade in dust radius.
Or according to Eq.~\ref{eq:binLength} this means that our reference decade fraction is $d_r=0.5$.
The target decade fraction shall be $d_t$.
First we define the helper function
\begin{equation}
    H(a_{0},N_{0},\beta,d,a) = \frac{C(a_{0},N_{0},\beta)}{(1-\beta)}   a^{1-\beta} \left( 10^{d(1-\beta)} - 1 \right) \quad,
\end{equation}
which allows us to calculate the integral, $I$, of Eq.~\ref{eq:colDenFit} for one dust bin:
\begin{align}
    I(a_{0,1},N_{0,1},\beta_1,a_{0,2},N_{0,2},\beta_2,d,a) &= H(a_{0,1},N_{0,1},\beta_1,d,a) \nonumber\\
    &+ H(a_{0,2},N_{0,2},\beta_2,d,a) \,.
\end{align}
From this we can now calculate the scaling constant, $S$, for a bin that was initially covered the decade fraction $d_r$ to one that covers the decade fraction $d_t$.
This scaling constant can be written as
\begin{align}
    S(a_{0,1},N_{0,1},\beta_1,&a_{0,2},N_{0,2},\beta_2,\mathbf{d_r},\mathbf{d_t},a) = \nonumber\\ 
    &\frac{I(a_{0,1},N_{0,1},\beta_1,a_{0,2},N_{0,2},\beta_2,\mathbf{d_t},a)}{I(a_{0,1},N_{0,1},\beta_1,a_{0,2},N_{0,2},\beta_2,\mathbf{d_r},a)} \,.
\end{align}
The column density of the new bin length, $N_{col}(a,d_t)$, can now be calculated from the column density given by the EDCM, $N_{col}(a,d_r)$:
\begin{align}
    N_{col}(a,\mathbf{d_t}) &= N_{col}(a,\mathbf{d_r}) \times \nonumber\\
    &S(a_{0,1},N_{0,1},\beta_1,a_{0,2},N_{0,2},\beta_2,\mathbf{d_r},\mathbf{d_t},a) \,.
\end{align}
Though we have kept all equations general, we remind the reader that currently $d_r=0.5$.

%%%%%%%%%%%%%%%%%%%%%%%%%%%%%%%%%%%%%%%%%%%%%%%%%%%%%%%%%%%%%%%%%%%%%%%%%%
%%%%%%%%%%%%%%%%%%%%%%%%%%%%%%%%%%%%%%%%%%%%%%%%%%%%%%%%%%%%%%%%%%%%%%%%%%
%%%%%%%%%%%%%%%%%%%%%%%%%%%%%%%%%%%%%%%%%%%%%%%%%%%%%%%%%%%%%%%%%%%%%%%%%%
%%%%%%%%%%%%%%%%%%%%%%%%%%%%%%%%%%%%%%%%%%%%%%%%%%%%%%%%%%%%%%%%%%%%%%%%%%
%%%%%%%%%%%%%%%%%%%%%%%%%%%%%%%%%%%%%%%%%%%%%%%%%%%%%%%%%%%%%%%%%%%%%%%%%%
%%%%%%%%%%%%%%%%%%%%%%%%%%%%%%%%%%%%%%%%%%%%%%%%%%%%%%%%%%%%%%%%%%%%%%%%%%
\section{Discussion \& Conclusions}\label{sec:discussion}
In this work we have considered a new approach to determine the dust environment of a yet unknown comet as the one to be intercepted by ESA's Comet Interceptor mission.
We have argued that it is not clear -- a priory -- how a best, nominal, or worst case scenario can be defined because of the non-linear interdependence of the parameters in any dust coma model.

We have therefore taken the opposite approach of defining ranges for all input parameters to dust coma model and identifying all possible parameter combinations.
The collection of all these combinations provide a set of coma environments which can then be analysed statistically to determine e.g. a median case.

The results we have presented show that the range of possible dust densities which can be encountered by a spacecraft span many orders of magnitude.
This large uncertainty is dominated by the unknowns of the parameters of the nucleus, the dust particles, and the coma.
As long as the uncertainties in these parameters remains large so does the uncertainty of the predicted dust environment.
But as we learn more about a target, these uncertainties can be successively diminished.

We should also point out that even though it is possible to define a clear median case obviously does not imply that it is the most probably case.
In this work we have made no attempt to quantify which input parameters are more likely than others.
The EDCM assumes that each parameter is equally likely. 
Thus the model is agnostic as to the likelihood of any parameter taking a specific value and assigns it an equal probability.
We should therefore not expect the median but rather any value within a certain confidence interval around it.

Our model shows that the brightness Af$\rho$ \citep[as defined in][where A is the bond albedo, f is the filling factor of grains within the field of view, and $\rho$ is the radius of the field of view]{AHearn1984AJ} alone is a poor indicator of the dust production rate.
In combination with a known dust size distribution the expected uncertainty can be reduced to $\sim$1 order of magnitude.

The parameters most predictive of the column density along the spacecraft fly-by are a combination of Af$\rho$, the power law exponent, and the dust production rate.
For small particles Af$\rho$ in combination with either the power law exponent or the dust production rate is the most predictive.
In contrast, for large particles the power law exponent in combination with either Af$\rho$ or the dust production rate is the most predictive.

Unfortunately, only Af$\rho$ (or an equivalently measured brightness in flux or magnitudes) is an observable. 
The size distributions and dust production rates are always model dependent.
Any assessment of the coma environment, therefore, boils down to what range of model parameters are possible to match a given Af$\rho$. 

A somewhat independent constraint on size distribution can be achieved using Finson-Probstein modelling of the dust tails \citep[e.g.][]{Agarwal2010Icar}. 
But even this size distribution is a model-dependent interpretation rather than a direct observational measurement.
Thus, because only Af$\rho$ is available as a direct measurement of the three most predictive parameters of the dust abundance in the coma, the only remaining approach is the one we have outlined in this work.
For a certain Af$\rho$ one can assess what comet/coma properties are consistent with that measurement.

For typical encounter velocities we have shown that the angle of impact of dust particles onto the spacecraft are dominated by the spacecraft motion.
This means that shielding of the spacecraft can be concentrated to the ram-side of the spacecraft.
Though there might still be implications for attitude control which need to be checked separately.

Finally, we provide all results to the public in easy to read formats\footnote{\url{https://www.doi.org/10.5281/zenodo.6906815}}.
Should the standard results we present here not satisfy a readers need we have provided some scaling relationships in Sec.~\ref{sec:scalingResults} to adapt our results.

\section*{Acknowledgements}
We acknowledge the support from the International Space Science Institute (ISSI) through the team "Closing The Gap Between Ground Based And In-Situ Observations Of Cometary Dust Activity: Investigating Comet 67P To Gain A Deeper Understanding Of Other Comets".\\
VDC acknowledges the support by the Italian Space Agency (ASI) within the ASI-INAF agreements I/032/05/0, I/024/12/0 and 2020-4-HH.0.

\bibliography{main}{}
\bibliographystyle{aasjournal}

%\bibliographystyle{alpha}
%\bibliography{sample}

\end{document}